# Pure Nash Equilibria: Hard and Easy Games

**Georg Gottlob**                                                    GOTTLOB@DBAI.TUWIEN.AC.AT
*Information Systems Department,*
*Technische Universität Wien,*
*A-1040 Wien, Austria*

**Gianluigi Greco**                                                    GGRECO@MAT.UNICAL.IT
*Dipartimento di Matematica,*
*Università della Calabria,*
*I-87030 Rende, Italy*

**Francesco Scarcello**                                                    SCARCELLO@DEIS.UNICAL.IT
*DEIS,*
*Università della Calabria,*
*I-87030 Rende, Italy*

## Abstract

We investigate complexity issues related to pure Nash equilibria of strategic games. We show that, even in very restrictive settings, determining whether a game has a pure Nash Equilibrium is NP-hard, while deciding whether a game has a strong Nash equilibrium is $\Sigma_2^P$-complete. We then study practically relevant restrictions that lower the complexity. In particular, we are interested in quantitative and qualitative restrictions of the way each player's payoff depends on moves of other players. We say that a game has small neighborhood if the utility function for each player depends only on (the actions of) a logarithmically small number of other players. The dependency structure of a game $\mathcal{G}$ can be expressed by a graph $G(\mathcal{G})$ or by a hypergraph $H(\mathcal{G})$. By relating Nash equilibrium problems to constraint satisfaction problems (CSPs), we show that if $\mathcal{G}$ has small neighborhood and if $H(\mathcal{G})$ has bounded hypertree width (or if $G(\mathcal{G})$ has bounded treewidth), then finding pure Nash and Pareto equilibria is feasible in polynomial time. If the game is graphical, then these problems are LOGCFL-complete and thus in the class $NC_2$ of highly parallelizable problems.

## 1. Introduction and Overview of Results

The theory of strategic games and Nash equilibria has important applications in economics and decision making (Nash, 1951; Aumann, 1985). Determining whether Nash equilibria exist, and effectively computing them, are relevant problems that have attracted much research in computer science (e.g. Deng, Papadimitriou, & Safra, 2002; McKelvey & McLennan, 1996; Koller, Megiddo, & von Stengel, 1996). Most work has been dedicated to complexity issues related to *mixed equilibria* of games with *mixed strategies*, where the player's choices are not deterministic and are regulated by probability distributions. In that context, the existence of a Nash equilibrium is guaranteed by Nash's famous theorem (Nash, 1951), but it is currently open whether such an equilibrium can be computed in polynomial time (cf., Papadimitriou, 2001). First results on the computational complexity for a two-person game have been presented by Gilboa and Zemel (1989), while extensions to more general types of games have been provided by Megiddo and Papadimitriou (1991), and by Papadimitriou (1994b). A recent paper of Conitzer and Sandholm (2003b) also proved the NP-hardness of determining whether Nash equilibria with certain natural properties exist.

In the present paper, we are not dealing with mixed strategies, but rather investigate the complexity of deciding whether there exists a Nash equilibrium in the case of *pure strategies*, where each player chooses to play an action in a deterministic, non-aleatory manner. Nash equilibria for pure strategies are briefly referred to as *pure Nash equilibria*. Note that in the setting of pure strategies, a pure Nash equilibrium is not guaranteed to exist (see, for instance, Osborne & Rubinstein, 1994). Particular classes of games having pure Nash equilibria have been studied by Rosenthal (1973), Monderer and Shapley (1993), and by Fotakis et al. (2002). Recently, Fabrikant et al. (2004) renewed the interest in the class of games defined Rosenthal (1973), called *congestion games*, by showing that a pure Nash equilibrium can be computed in polynomial time in the symmetric network case, while the problem is PLS-complete (Johnson, Papadimitriou, & Yannakakis, 1998) in general.

Our goal is to study fundamental questions such as the *existence* of pure Nash, Pareto, and strong Nash equilibria, the *computation* of such equilibria, and to find arguably realistic restrictions under which these problems become tractable. Throughout the paper, Pareto and strong Nash equilibria are considered only in the setting of pure strategies.

While pure strategies are conceptually simpler than mixed strategies, the associated computational problems appear to be harder. In fact, we show that even if severe restrictions are imposed on the set of allowed strategies, determining whether a game has a pure Nash or Pareto Equilibrium is NP-complete, while deciding whether a game has a strong Nash equilibrium is even $\Sigma_2^P$-complete. However, by jointly applying suitable pairs of more realistic restrictions, we obtain settings of practical interest in which the complexity of the above problems is drastically reduced. In particular, determining the existence of a pure Nash equilibrium and computing such an equilibrium will be feasible in polynomial time and we will show that, in certain cases, these problems are even complete for the very low complexity class LOGCFL, which means that these problems are essentially as easy as the membership problem for context-free languages, and are thus highly parallelizable (in $NC_2$).

In the setting of pure strategies, to which we will restrict our attention in the rest of this paper, a finite strategic game is one in which each player has a finite set of possible actions, from which she chooses an action once and for all, independently of the actual choices of the other players. The choices of all players can thus be thought to be made simultaneously. The choice of an action by a player is referred to as the player's *strategy*. It is assumed that each player has perfect knowledge over all possible actions and over the possible strategies of all players. A *global strategy*, also called *profile* in the literature, consists of a tuple containing a strategy for each player. Each player has a polynomial-time computable real valued *utility function*, which allows her to assess her subjective utility of each possible global strategy (global strategies with higher utility are better). A pure *Nash equilibrium* (Nash, 1951) is a global strategy in which no player can improve her utility by changing her action (while the actions of all other players remain unchanged). A *strong Nash equilibrium* (Aumann, 1959) is a pure Nash equilibrium where no change of strategies of whatever coalition (i.e., group of players) can simultaneously increase the utility for all players in the coalition. A pure Nash equilibrium is *Pareto optimal* (e.g. Maskin, 1985) if the game admits no other pure Nash equilibrium for which each player has a strictly higher utility. A Pareto-optimal Nash equilibrium is also called a Pareto Nash Equilibrium.

Before describing our complexity results, let us discuss various parameters and features that will lead to restricted versions of strategic games. We consider restrictions of strategic games which impose quantitative and/or qualitative limitations on how the payoffs of an agent (and hence her decisions) may be influenced by the other agents.

The *set of neighbors Neigh(p)* of a player is the set of other players who potentially matter w.r.t. $p$'s utility function. Thus, whenever a player $q \neq p$ is not in $Neigh(p)$ then $p$'s utility function does not directly depend on the actions of $q$. We assume that each game is equipped with a polynomial-time computable function *Neigh* with the above property.[1] The player neighborhood relationship, typically represented as a graph (or a hypergraph), is the central notion in *graphical games* (Koller & Milch, 2001; Kearns, Littman, & Singh, 2001b), as we will see in more detail in the next section.

A first idea towards the identification of tractable classes of games is to restrict the cardinality of $Neigh(p)$ for all players $p$. For instance, consider a set of companies in a market. Each company has usually a limited number of other market players on which it bases its strategic decisions. These relevant players are usually known and constitute the neighbors of the company in our setting. However, note that even in this case the game outcome still depends on the interaction of all players, though possibly in an indirect way. Indeed, the choice of a company influences the choice of its competitors, and hence, in turn, the choice of competitors of its competitors, and so on. In this more general setting, a number of real-world cases can be modeled in a very natural way. We can thus define the following notion of limited neighborhood:

**Bounded Neighborhood:** Let $k > 0$ be a fixed constant. A strategic game with associated neighborhood function *Neigh* has *k-bounded neighborhood* if, for each player $p$, $|Neigh(p)| \leq k$.

While in some setting the bounded neighborhood assumption is realistic, in other settings the constant bound appears to be too harsh an imposition. It is much more realistic and appealing to relax this constraint and consider a *logarithmic* bound rather than a constant bound on the number of neighbors.

**Small Neighborhood:** For a game $\mathcal{G}$ denote by $P(\mathcal{G})$ the set of its players, by $Act(p)$ the set of possible actions of a player $p$, and by $||\mathcal{G}||$ the total size of the description of a game $\mathcal{G}$ (i.e., the input size $n$). Furthermore, let $maxNeigh(\mathcal{G}) = max_{p \in P(\mathcal{G})}|Neigh(p)|$ and $maxAct(\mathcal{G}) = max_{p \in P(\mathcal{G})}|Act(p)|$.

A class of strategic games has *small neighborhood* if, for each game $\mathcal{G}$ in this class,

$$maxNeigh(\mathcal{G}) = O(\frac{\log ||\mathcal{G}||}{\log maxAct(\mathcal{G})})$$

Note the denominator $\log maxAct(\mathcal{G})$ in the above bound. Intuitively, we use this term to avoid "cheating" by trading actions for neighbors. Indeed, roughly speaking, player interactions may be reduced significantly by adding an exponential amount of additional actions. For any player, these fresh actions may encode all possible action configurations of some of her neighbors, yielding an equivalent game with less interaction, and possibly with fewer players, too. The denominator takes this into account.

---

1. Note that each game can be trivially represented in this setting, possibly setting $Neigh(p)$ to be the set of all players, for each player $p$. In most cases, however, one will be able to provide a much better neighborhood function.

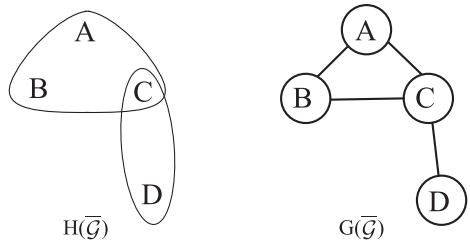

Figure 1: Dependency hypergraph and dependency graph for the game $\overline{\mathcal{G}}$.

In other terms, a class of games has small neighborhood if there is a constant $c$ such that for all but finitely many pairs $(\mathcal{G}, p)$ of games and players, $|Neigh(p)| < c \times (\frac{\log ||\mathcal{G}||}{\log |Act(p)|})$.

The related notion $i(\mathcal{G})$ of *intricacy* of a game is defined by:

$$i(\mathcal{G}) = \frac{maxNeigh(\mathcal{G}) \times \log maxAct(\mathcal{G})}{\log ||\mathcal{G}||}.$$

It is clear that a class of games has small neighborhood if and only if the intricacy of all games in it is bounded by some constant.

Obviously, bounded neighborhood implies small neighborhood, but not vice-versa. We believe that a very large number of important (classes of) games in economics have the small neighborhood property.

In addition to the quantitative aspect of the size of the neighborhood (and of the neighborhood actions), we are also interested in qualitative aspects of mutual strategic influence. Following Kearns et al. (2001b), for a game $\mathcal{G}$ with a set $P$ of players, we define the *strategic dependency graph* as the undirected graph $G(\mathcal{G})$ having $P$ as its set of vertices and $\{\{p, q\} \mid q \in P \wedge p \in Neigh(q)\}$ as its set of edges. Moreover, we define the *strategic dependency hypergraph* $H(\mathcal{G})$, whose vertices are the players $P$ and whose set of hyperedges is $\{\{p\} \cup Neigh(p) \mid p \in P\}$. For instance, consider a game $\overline{\mathcal{G}}$ over players $A, B, C$, and $D$ such that $Neigh(A) = \{B, C\}$, $Neigh(B) = \{A, C\}$, $Neigh(C) = \{A, B\}$, and $Neigh(D) = \{C\}$. Figure 1 shows the dependency graph and the dependency hypergraph associated with $\overline{\mathcal{G}}$.

We consider the following classes of structurally restricted games:

**Acyclic-Graph Games:** Games $\mathcal{G}$ for which $G(\mathcal{G})$ is acyclic.

**Acyclic-Hypergraph Games:** Games $\mathcal{G}$ for which $H(\mathcal{G})$ is acyclic. Note that there are several definitions of hypergraph acyclicity (Fagin, 1983). Here we refer to the broadest (i.e., the most general) one, also known as $\alpha$-acyclicity (Fagin, 1983; Beeri, Fagin, Maier, & Yannakakis, 1983) (see Section 2).

Each acyclic-graph game is also an acyclic-hypergraph game, but not vice-versa. As an extreme example, let $\mathcal{G}$ be a game with player set $P$ in which the utility of each action for each player depends on all other players. Then $G(\mathcal{G})$ is a clique of size $|P|$ while $H(\mathcal{G})$ is the trivially acyclic hypergraph having the only hyperedge $\{P\}$.

For strategic games, both the acyclic graph and the acyclic hypergraph assumptions are very severe restrictions, which are rather unlikely to apply in practical contexts. However, there are important generalizations that appear to be much more realistic for practical applications. These concepts are bounded treewidth (Robertson & Seymour, 1986) and bounded hypertree width (Gottlob, Leone, & Scarcello, 2002b) (see also Section 5), which are suitable measures of the degree of cyclicity of a graph and of a hypergraph, respectively. In particular, each acyclic graph (hypergraph) has treewidth (hypertree width) $\leq 1$. It was argued that an impressive number of "real-life" graphs have a very low treewidth (Downey & Fellows, 1995). Hypertree width in turn was fruitfully applied in the context of database queries (Gottlob et al., 2002b) and constraint satisfaction problems (Gottlob, Leone, & Scarcello, 2000). Formal definitions are given in Section 5. Note that both computing the treewidth of a graph and the hypertree width of a hypergraph are NP-hard problems. However, for each (fixed) constant $k$, it can be checked in polynomial time whether a graph has treewidth $k$ (Bodlaender, 1997) and whether a hypergraph has hypertree width $k$ (Gottlob et al., 2002b). We have, for each constant $k$, the following restricted classes of games:

**Games of treewidth bounded by $k$:** The games $\mathcal{G}$ such that the treewidth of $G(\mathcal{G})$ is $\leq k$.

**Games of hypertree width bounded by $k$:** The games $\mathcal{G}$ such that the hypertree width of $H(\mathcal{G})$ is $\leq k$.

In the context of complexity and efficiency studies, it is very important to make clear how an input (in our case, a multiplayer game) is represented. We say that a game is in *general form* if the sets of players and actions are given in extensional form and if the neighborhood and utility functions are polynomially computable functions. Unless otherwise stated, we always assume that games are given in general form. For classes of games having particular properties, some alternative representations have been used by various authors. For instance, in game theory literature, the set of utility functions is often represented through a single table (or matrix) having an entry for *each* combination of players' actions containing, for each player $p$, the evaluation of her utility function for that particular combination. This representation is said to be in *standard normal form (SNF)* (see, for instance, Osborne & Rubinstein, 1994; Owen, 1982). Note that, if there are many players, this representation may be very space consuming, particularly if some players are not interested in all other players, but only in some subset of them. Moreover, in this case, the monolithic utility table in SNF obscures much of the structure that is present in real-world games (Koller & Milch, 2001). In fact, in the context of games with restricted players interactions, the most used representation is the *graphical normal form (GNF)*. In GNF games, also known as *graphical games* (Kearns, Littman, & Singh, 2001a; Kearns et al., 2001b; Kearns & Mansour, 2002; Vickrey, 2002), the utility function for each player $p$ is given by a table that displays $p$'s utility as a function of all possible combined strategies of $p$ and $p$'s neighbors, but not of other players irrelevant to $p$. Therefore, for large population games (modeling for instance agent interactions over the internet), the SNF is practically unfeasible, while the more succinct graphical normal form works very well, and is actually a more natural representation.

**Main results.** The main results of this paper are summarized as follows:

- Determining whether a strategic game has a pure Nash equilibrium is NP-complete and remains NP-complete even for following two restricted cases:

  - Games in graphical normal form (GNF) having bounded neighborhood (Theorem 3.1).
  - Acyclic-graph games, and acyclic-hypergraph games (Theorem 3.2).

  The same results hold for Pareto Nash equilibria for pure strategies.

- Determining whether a strategic game has a strong Nash equilibrium is $\Sigma_2^p$-complete and thus at the second level of the Polynomial Hierarchy (Theorem 3.7 and Theorem 3.8). The proof of this theorem gives us a fresh game-theoretic view of the class $\Sigma_2^P$ as the class of problems whose positive instances are characterized by a coalition of players who cooperate to provide an equilibrium, and win against any other disjoint coalition, which fails in trying to improve the utility for all of its players. E.g., in the case of $\Sigma_2$ quantified Boolean formulas, the former coalition consists of the existentially quantified variables, and the latter of the universally quantified ones.

- The pure Nash, Pareto and strong equilibrium existence and computations problems are feasible in logarithmic space for games in standard normal form (Theorem 4.1).

- The pure Nash equilibrium existence and computation problems are tractable for games (in whatever representation) that simultaneously have small neighborhood and bounded hypertree width (Theorem 5.3). Observe that each of the two joint restrictions, small neighborhood and bounded hypertree width, is weaker than the restrictions of bounded neighborhood and acyclicity, respectively, of which each by itself does not guarantee tractability. Thus, in order to obtain tractability, instead of strengthening a single restriction, we combined two weaker restrictions. While we think that each of the two strong restrictions is unrealistic, we believe that for many natural games the combined weaker restrictions do apply. In order to prove the tractability result, we establish a relationship between strategic games and the well-known finite domain constraint satisfaction problem (CSP), much studied in the AI and OR literature (e.g. Vardi, 2000; Gottlob et al., 2000). Let us point out that also Vickrey and Koller (2002) recently exploited a mapping to CSP for the different problem of finding approximate mixed equilibria in graphical games. We show that each (general, not necessarily GNF) strategic game $\mathcal{G}$ can be translated into a CSP instance having the same hypertree width as $\mathcal{G}$, and whose feasible solutions exactly correspond to the Nash equilibria of the game. Then, we are able to prove that $\mathcal{G}$ is equivalent to an *acyclic* constraint satisfaction problem of size $||\mathcal{G}||^{O(i(\mathcal{G}) \times hw(\mathcal{G}))}$, where $i(\mathcal{G})$ is the intricacy of $\mathcal{G}$ and $hw(G)$ is the hypertree width of its strategic dependency hypergraph. Acyclic CSPs, in turn, are well-known to be solvable in polynomial time.

- Exploiting the same relationship with CSPs, we prove that the Nash-equilibrium existence and computation problems are tractable for games in graphical normal form (GNF) having bounded hypertree width (Theorem 5.3), regardless of the game intricacy, i.e., even for unbounded neighborhood.

- We show that if a strategic game has bounded treewidth, then it also has bounded hypertree width (Theorem 5.7). Note that this is a novel result on the relationship between these two measures of the degree of cyclicity, since earlier works on similar subjects dealt with either the primal or the dual graph of a given hypergraph, rather than with a dependency graph, as we do in the present paper, focused on games. Combined with the two previous points, this entails that the Nash-equilibrium existence and computation problems are tractable for games that simultaneously have small neighborhood and bounded treewidth, and for GNF games having bounded treewidth (Corollary 5.9).

- In all above cases where a pure Nash Equilibrium can be computed in polynomial time, also a Pareto Nash equilibrium can be computed in polynomial time (Theorem 4.6 and Corollary 5.4).

- These tractability results partially extend to strong Nash equilibria. Indeed, the checking problem becomes feasible in polynomial time for acyclic-hypergraph games in GNF. However, even in such simple cases, deciding whether a game has strong Nash equilibria is NP-complete, and thus still untractable (Theorem 4.8).

- We go a bit further, by determining the precise complexity of games with acyclic (or even bounded width) interactions among players: In case a game is given in GNF, the problem of determining a pure Nash equilibrium of a game of bounded hypertree-width (or bounded treewidth) is LOGCFL-complete and thus in the parallel complexity class $NC_2$ (Theorem 6.1). Membership in LOGCFL follows from the membership of bounded hypertree-width CSPs in LOGCFL (Gottlob, Leone, & Scarcello, 2001). Hardness for LOGCFL is shown by transforming (logspace uniform families of) semi-unbounded circuits of logarithmic depth together with their inputs into strategic games, such that the game admits a Nash equilibrium if and only if the circuit outputs one on the given input.

Figure 2 summarizes our results on the existence of pure Nash equilibria. While various authors have dealt with the complexity of Nash equilibria (e.g. Gilboa & Zemel, 1989; Papadimitriou, 1994b; Koller & Megiddo, 1992, 1996; Conitzer & Sandholm, 2003b), most investigations were dedicated to mixed equilibria and — to the best of our knowledge — all complexity results in the present paper are novel. We are not aware of any other work considering the quantitative and structural restrictions on pure games studied here. Note that tree-structured games were first considered by Kearns et al. (2001b) in the context of mixed equilibria. It turned out that, for such games, suitable approximation of (mixed) Nash equilibria can be computed in polynomial time. In our future work, we would like to extend our tractability results even to this setting. We are not aware of any work by others on the parallel complexity of equilibria problems. We believe that our present work contributes to the understanding of pure Nash equilibria and proposes appealing and realistic restrictions under which the main computation problems associated with such equilibria are tractable.

The rest of the paper is organized as follows. In Section 2 we introduce the basic notions of games and Nash equilibria that are studied in the paper, and we describe how games may be represented. In Section 3 we thoroughly study the computational complexity of deciding the existence of pure Nash, Pareto and strong equilibria. In Section 4 we identify tractable classes of games, and in Section 5 we extend our tractability results to larger class of games, where the interaction among players has a bounded degree of cyclicity. In Section 6 we improve the results on the polynomial tractability of easy games, providing the precise computational complexity for games with acyclic (or bounded width) interactions among players. Finally, in Section 7, we draw our conclusions, and we discuss possible further research and related works.

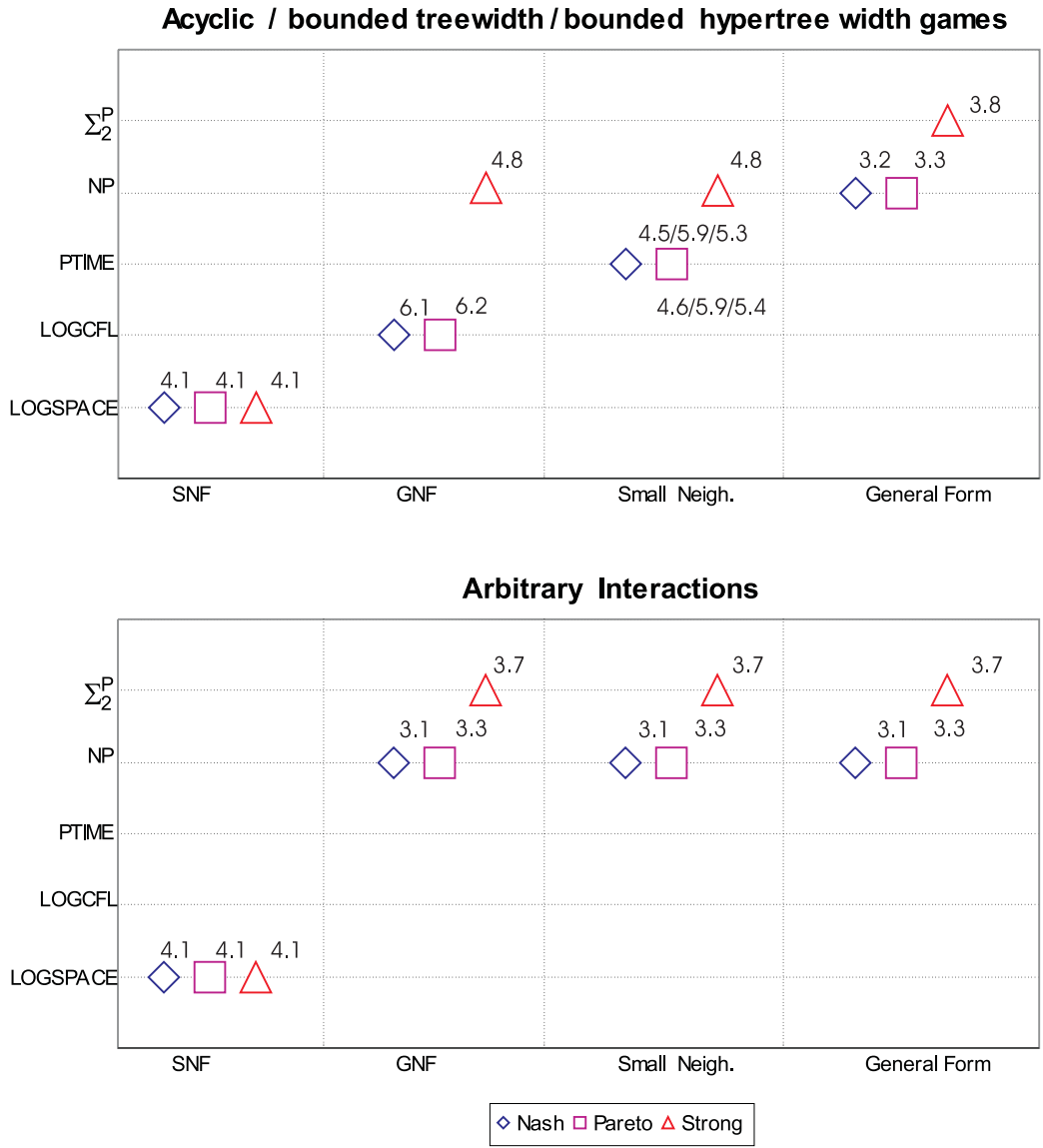

Figure 2: Complexity of deciding existence of pure Nash equilibria for games in GNF — numbers indicate theorems where the corresponding results are proved.

## 2. Games and Nash Equilibria

A *game* $\mathcal{G}$ is a tuple $\langle P, Neigh, Act, U \rangle$, where $P$ is a non-empty set of distinct players and $Neigh : P \longrightarrow 2^P$ is a function such that for each $p \in P$, $Neigh(p) \subseteq P - \{p\}$ contains all neighbors of $p$, $Act : P \longrightarrow A$ is a function returning for each player $p$ a set of possible actions $Act(p)$, and $U$ associates a utility function $u_p : Act(p) \times_{j \in Neigh(p)} Act(j) \to \Re$ to each player $p$.

Note that, in general, the players interests are not symmetric. Thus, it may happen that, for a pair of players $p_1, p_2 \in P$, $p_1 \in Neigh(p_2)$ but $p_2 \notin Neigh(p_1)$.

For a player $p$, $p_a$ denotes her choice to play the action $a \in Act(p)$. Each possible $p_a$ is called a *strategy* for $p$, and the set of all strategies for $p$ is denoted by $St(p)$.[2]

For a non-empty set of players $P' \subseteq P$, a *combined strategy* for $P'$ is a set containing exactly one strategy for each player in $P'$. $St(P')$ denotes the set of all combined strategies for the players in $P'$.

A combined strategy (also, *profile*) $\mathbf{x}$ is called *global* if all players contribute to it, that is, if $P' = P$. The global strategies are the possible outcomes of the game.

A set of players $K \subseteq P$ is often called a *coalition*. Let $\mathbf{x}$ be a global strategy, $K$ a coalition, and $\mathbf{y}$ a combined strategy for $K$. Then, we denote by $\mathbf{x}_{-K}[\mathbf{y}]$ the global strategy where, for each player $p \in K$, her individual strategy $p_a \in \mathbf{x}$ is replaced by her individual strategy $p_b \in \mathbf{y}$. If $K$ is a singleton $\{p\}$, we will simply write $\mathbf{x}_{-p}[\mathbf{y}]$.

Let $\mathbf{x}$ be a global strategy, $p$ a player, and $u_p$ the utility function of $p$. Then, with a small abuse of notation, $u_p(\mathbf{x})$ will denote the output of $u_p$ on the projection of $\mathbf{x}$ to the domain of $u_p$, i.e., the output of the function $u_p$ applied to the actions played by $p$ and her neighbors according to the strategy $\mathbf{x}$.

In the context of complexity and efficiency studies it is very important to make clear how an input (in our case, a multiplayer game) is represented.

**General Form:** A game is in *general form* if the sets of players and actions are given in extensional form, while the neighborhood and utility functions are given intentionally, e.g., through encodings of deterministic Turing transducers. More precisely, we are interested in classes of games such that the computation time of the neighborhood and utility functions is globally bounded by some polynomial. Let us denote by $\mathcal{C}_k$ the class of all games $\mathcal{G}$ in general form whose neighborhood and utility functions are computable in time $O(n^k)$, where $n = |\mathcal{G}|$.

For the sake of presentation, we assume hereafter $\bar{k} \geq 1$ to be any such a fixed global bound. Moreover, unless otherwise stated, when we speak of a "general game $\mathcal{G}$" (or we omit any specification at all) we mean a game $\mathcal{G} \in \mathcal{C}_{\bar{k}}$.

The following more restrictive classes of (representations of) games have been used by many authors.

**Standard Normal Form (SNF):** A game with set $P$ of players is in *standard normal form (SNF)* if its utility functions are explicitly represented by a single table or matrix having an entry (or cell) for *each* global strategy $\mathbf{x}$, displaying a list containing for each player $p$, $p$'s payoff $u_p(\mathbf{x})$ w.r.t. $\mathbf{x}$. (Equivalently, we may describe the utilities by $|P|$ such tables, where the $i$-th table describes just the payoff of player $i$.) This is a representation of utility functions often assumed in the literature (see, for instance, Osborne & Rubinstein, 1994; Owen, 1982). Observe that, in the general case, even if an utility function is polynomially computable, writing it down in form of a table may require exponential space.

---

2. Note the technical distinction between actions and strategies: an action is an element of the form $a$, while a strategy is an element of the form $p_a$, i.e., it is an action chosen by a player. This helps in technical proofs, since a strategy singles out both a player and her choice.

**Graphical Normal Form (GNF):**  A game with a set $P$ of players is in *graphical normal form (GNF)* if the utility function for each player $p$ is represented by a separate table containing a cell for each combined strategy $\mathbf{x} \in St(Neigh(p) \cup \{p\})$ of the $p$'s set of neighbors $Neigh(p) \cup \{p\}$, displaying $p$'s payoff $u_p(\mathbf{x})$ w.r.t. $\mathbf{x}$. A game in GNF is illustrated in Example 2.1. The GNF representation has been adopted in several recent papers that study games with a large number of players, where the utility function of each player depends directly only on the strategies of those (possibly few) players she is interested in (e.g. Kearns et al., 2001a, 2001b; Kearns & Mansour, 2002; Vickrey, 2002). Note that GNF may lead to an exponentially more succinct game representation than SNF. Notwithstanding, the SNF is often used in the literature, mostly because, historically, the first investigations focused on two-player games. Moreover, games in GNF are often referred to as *graphical games.* We prefer to use the phrasing *games in graphical normal form*, because this makes clear that we are addressing representational issues.

The following example, to which we will refer throughout the paper, should sound familiar to everyone, as it is a generalization of the well known two-person game "battle of sexes".

**Example 2.1 (FRIENDS)** Let us consider the game FRIENDS, that is played by a group of persons that have to plan their evening happenings. The players are *George* (short: $G$), *Pauline* ($P$), *Frank* ($F$), *Robert* ($R$), and *Mary* ($M$). Each of them have to decide to go either to see a *movie* ($m$) or to see an *opera* ($o$). However, preferences concern not only the particular option ($m$ or $o$) to be chosen, but usually also the persons to join for the evening (possibly, depending on the movie or opera choice). For instance, we assume that Frank is interested in joining Pauline and Robert. He would like to join both of them. However, Pauline is an expert of movies and Robert is an expert of operas. Thus, if it is not possible to go out all together, he prefers to go to the cinema with Pauline and to the opera with Robert. Pauline would like to stay with Frank, and she prefers the movies. Robert does not like Frank because he speaks too much and, as we know, he prefers the opera. Mary, too, likes operas and would like to go to the opera with Robert. Finally, George is the matchmaker of the group: He has no personal preferences but would like that Frank and Pauline stay together for the evening, best if they go to the cinema. All the utility functions associated with this game are shown in Figure 3, where we denote the fact that a player $X$ chooses an action $a$ by $X_a$, e.g., $F_m$ denotes the strategy where Frank chooses to play the action $m$. □

Let us now formally define the main concepts of equilibria to be further studied in this paper.

**Definition 2.2** Let $\mathcal{G} = \langle P, Neigh, A, U \rangle$ be a game. Then,

| $F$ | $P_mR_m$ | $P_mR_o$ | $P_oR_m$ | $P_oR_o$ |
|---|---|---|---|---|
| $m$ | 2 | 2 | 1 | 0 |
| $o$ | 0 | 2 | 1 | 2 |

| $G$ | $P_mF_m$ | $P_mF_o$ | $P_oF_m$ | $P_oF_o$ |
|---|---|---|---|---|
| $m$ | 2 | 0 | 0 | 1 |
| $o$ | 2 | 0 | 0 | 1 |

| $R$ | $F_m$ | $F_o$ |
|---|---|---|
| $m$ | 0 | 1 |
| $o$ | 2 | 0 |

| $P$ | $F_m$ | $F_o$ |
|---|---|---|
| $m$ | 2 | 0 |
| $o$ | 0 | 1 |

| $M$ | $R_m$ | $R_o$ |
|---|---|---|
| $m$ | 1 | 0 |
| $o$ | 0 | 2 |

Figure 3: Utility functions for FRIENDS in GNF

- a global strategy $\mathbf{x}$ is a pure Nash Equilibrium for $\mathcal{G}$ if, for every player $p \in P$, $\nexists p_a \in St(p)$ such that $u_p(\mathbf{x}) < u_p(\mathbf{x}_{-p}[p_a])$;

- a global strategy $\mathbf{x}$ is a pure strong Nash Equilibrium for $\mathcal{G}$ if, $\forall K \subseteq P$, $\forall y \in St(K)$, $\exists p \in K$ such that $u_p(\mathbf{x}) \geq u_p(\mathbf{x}_{-K}[y])$ or, equivalently, if $\forall K \subseteq P$, $\nexists y \in St(K)$ such that, $\forall p \in K$, $u_p(\mathbf{x}) < u_p(\mathbf{x}_{-K}[y])$;

- a pure Nash equilibrium $\mathbf{x}$ is a pure Pareto Nash Equilibrium for $\mathcal{G}$ if there does not exist a pure Nash equilibrium $\mathbf{y}$ for $\mathcal{G}$ such that, $\forall p \in P$, $u_p(\mathbf{x}) < u_p(\mathbf{y})$.[3]

The sets of pure Nash, strong Nash, and Pareto Nash equilibria of $\mathcal{G}$ are denoted by $\mathcal{NE}(\mathcal{G})$, $\mathcal{SNE}(\mathcal{G})$, and $\mathcal{PNE}(\mathcal{G})$, respectively. It is easy to see and well known that the following relationships hold among these notions of Nash equilibria: $\mathcal{SNE}(\mathcal{G}) \subseteq \mathcal{PNE}(\mathcal{G}) \subseteq \mathcal{NE}(\mathcal{G})$. Moreover, the existence of a Nash equilibrium does not imply the existence of a strong Nash equilibrium. However, if there exists a Nash equilibrium, then there exists also a Pareto Nash equilibrium.

**Example 2.3** The strategies $\{F_m, P_m, R_o, G_m, M_o\}$, $\{F_m, P_m, R_o, G_o, M_o\}$, $\{F_o, P_o, R_m, G_m, M_m\}$ and $\{F_o, P_o, R_m, G_o, M_m\}$ are the Nash equilibria of the FRIENDS game. For instance, consider the latter strategy, where all players get payoff 1. In this case, since $P$ plays opera and $R$ plays movie, $F$ cannot improve his payoff by changing from opera to movie. The same holds for $G$, while $R$, $P$, and $M$ would get the lower payoff 0, if they change their choices.

Moreover, note that the first two strategies above, namely $\{F_m, P_m, R_o, G_m, M_o\}$ and $\{F_m, P_m, R_o, G_o, M_o\}$, are the only Pareto Nash equilibria, as well as the strong Nash equilibria. Indeed, for these global strategies all players get their maximum payoff 2, and thus there is no way to improve their utilities. □

The interaction among players of $\mathcal{G}$ can be more generally represented by a hypergraph $H(\mathcal{G})$ whose vertices coincide with the players of $\mathcal{G}$ and whose set of (hyper)edges contains for each player $p$ a (hyper)edge $H(p) = \{p\} \cup Neigh(p)$, referred-to as the *characteristic edge of $p$*. Intuitively, characteristic edges correspond to utility functions.

---

3. Note that only pure strategies do matter in this definition, as there is no requirement with regard to how pure candidate equilibria compare to possible mixed equilibria.

A fundamental structural property of hypergraphs is *acyclicity*. Acyclic hypergraphs have been deeply investigated and have many equivalent characterizations (e.g. Beeri et al., 1983). We recall here that a hypergraph H is acyclic if and only if there is a *join tree* for H, that is, there is a tree $JT$ whose vertices are the edges of H and, whenever the same player $p$ occurs in two vertices $v_1$ and $v_2$, then $v_1$ and $v_2$ are connected in $JT$, and $p$ occurs in each vertex on the unique path linking $v_1$ and $v_2$ (see Figure 5 for a join tree of H(FRIENDS)). In other words, the set of vertices in which $p$ occurs induces a (connected) subtree of $JT$. We will refer to this condition as the *Connectedness Condition* of join trees (also known as *running intersection property*).

Another representation of the interaction among players is through the (undirected) dependency graph $G(\mathcal{G}) = (P, E)$, whose vertices coincide with the players of $\mathcal{G}$, and $\{p, q\} \in E$ if $p$ is a neighbor of $q$ (or vice versa). For completeness we observe that, even if most works on graphical games use this dependency graph, another natural choice is representing the game structure by a directed graph (also called influence graph), which takes into account the fact that payoffs of a player $p$ may depend on payoffs of a player $q$ and not vice versa, in general. Following Kearns et al. (2001b), in the present paper we use the undirected version because we are interested in identifying game structures that possibly allow us to compute efficiently Nash equilibria, and directed graphs do not help very much for this purpose. Let us give a hint of why this is the case, by thinking of a group of players $\bar{X} = \{X_1, \ldots, X_n\}$, each one having only one neighbor $C$, whose payoffs do not depend on any player in $\bar{X}$. Player $C$ has one neighbor $D$, who has $C$ has its only neighbor. Figure 4 shows a directed graph representing these player interactions. It is easy to design a game with these players such that, for some combined strategy **x** of the $\bar{X}$ players, there is no Nash equilibrium, while for some other combined strategy **x**′ of these players, there is a combined strategy **y** of $C$ and $D$ such that the union of **x**′ and **y** is a Nash equilibrium for the game. Therefore, as far as the possibility of reaching a Nash equilibrium is concerned, the choices of players in $\bar{X}$ depend on each other, on the way of playing of $C$, and transitively on player $D$, too. Observe that the undirected dependency graph represents in a succinct way, i.e., through their direct connections, such a mutual relationship among players. However, the direct graph does not model this kind of influence, as looking at this graph it seems that players in $\bar{X}$ should not worry about any other player in the game. In fact, exploiting the gadgets and the constructions described in this paper, it is easy to see that even simple games whose directed influence graphs are quasi-acyclic (i.e., they are acyclic, but for some trivial cycles like the one shown in Figure 4), are hard to deal with. Thus, apart from the well known easy acyclic cases, any reasonable generalization of the notion of direct acyclicity does not appear to be useful for identifying further tractable classes of games.

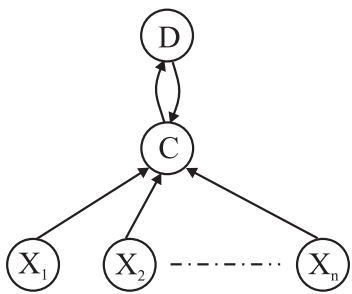

Figure 4: A directed graph representation of player interactions.

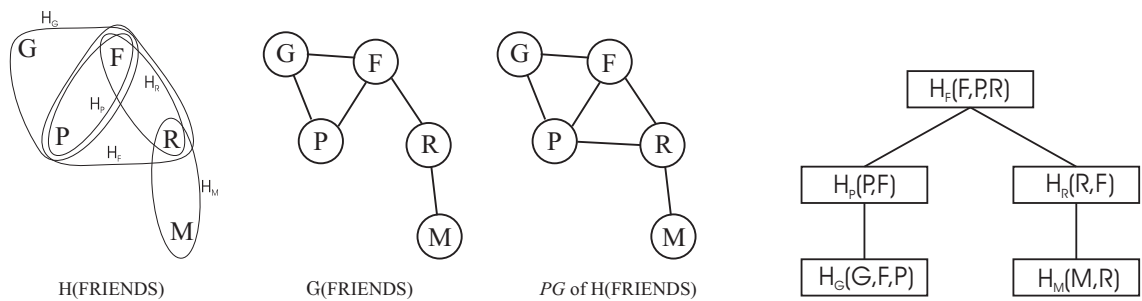

Figure 5: Hypergraph, dependency graph, and primal graph of the FRIENDS game. On the right, a join tree for H(FRIENDS).

Observe that the dependency graph G($\mathcal{G}$) is different from the so called *primal graph PG* of H($\mathcal{G}$), which contains an edge for all pairs of vertices that jointly occur in some hyperedge of H($G$). In general, G($\mathcal{G}$) is much simpler than $PG$. For instance, consider a game $\mathcal{G}$ with a player $p$ that depends on all other players $q_1, \ldots, q_n$, while these players are independent of each other (but possibly depend on $p$). Then, G($\mathcal{G}$) is a tree. However, the primal graph of H($\mathcal{G}$) is a clique with $n + 1$ vertices.

**Example 2.4** The hypergraph H(FRIENDS), the graph G(FRIENDS) and the primal graph of H(FRIENDS) are shown in Figure 5. Note that the dependency graph associated with the FRIENDS game is not acyclic, even though the associated hypergraph is acyclic (on the right, we also report a join tree for it). Moreover, note that the dependency graph differs from the primal graph, as player $P$ is not a neighbor of $R$ and viceversa.  □

## 3. Hard Games

In this section, we precisely characterize the complexity of deciding the existence of the different kinds of pure Nash equilibria (regular, Pareto, and strong). This way, we are able to identify the sources of complexity of such problems, in order to single out, in the following sections, some natural and practically relevant tractable cases.

Every game considered in this section is assumed to be either in general or in graphical normal form. Indeed, as we shall discuss in details in Section 4, for games in standard normal form, computing pure Nash equilibria is a tractable problem, since one can easily explore the (big) table representing the utility functions of all players, in order to detect the strategy of interest. Since this table is given in input, such a computation is trivially feasible in polynomial time.

We start by showing that deciding the existence of a pure Nash equilibrium is a difficult problem, even in a very restricted setting.

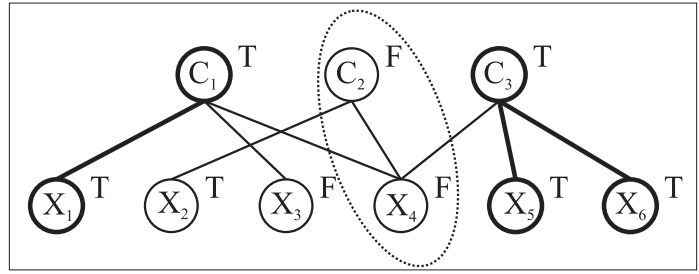

Figure 6: Schema of the reduction in the proof of Theorem 3.1.

**Theorem 3.1** *Deciding whether a game $\mathcal{G}$ has a pure Nash equilibrium is NP-complete. Hardness holds even if $\mathcal{G}$ is in GNF and has 3-bounded neighborhood, and the number of actions is fixed.*

**Proof.** *Membership.* We can decide that $\mathcal{NE}(\mathcal{G}) \neq \emptyset$ by guessing a global strategy $\mathbf{x}$ and verifying that $\mathbf{x}$ is a Nash equilibrium for $\mathcal{G}$. The latter task can be done in polynomial time. Indeed, for each player $p$ and for each action $a \in Act(p)$, we only have to check that choosing the strategy $p_a$ does not lead to an increment of $u_p$, and each of these tests is feasible in polynomial time.

*Hardness.* Recall that deciding whether a Boolean formula in conjunctive normal form $\Phi = c_1 \wedge \ldots \wedge c_m$ over the variables $X_1, \ldots, X_n$ is satisfiable, i.e., deciding whether there exists truth assignments to the variables making each clause $c_j$ true, is the well known NP-complete problem (CNF) SAT. Hardness holds even for its 3SAT restriction where each clause contains at most three distinct (possibly negated) variables, and each variable occurs in at most three clauses (Garey & Johnson, 1979). W.l.o.g, assume $\Phi$ contains at least one clause and one variable.

We define a GNF game $\mathcal{G}$ such that: The players are partitioned into two sets $P_v$ and $P_c$, corresponding to the variables and to the clauses of $\Phi$, respectively; for each player $c \in P_c$, $Neigh(c)$ is the set of the players corresponding to the variables in the clause $c$, and for each player $v \in P_v$, $Neigh(v)$ is the set of the players corresponding to the clauses in which $v$ occurs; $\{t, f, u\}$ is the set of possible actions for all the players, in which $t$ and $f$ can be interpreted as the truth values *true* and *false* for variables and clauses. Figure 6 shows the graph $\mathrm{G}(\mathcal{G})$ for the game $\mathcal{G}$ associated with the formula $(X_1 \vee X_3 \vee X_4) \wedge (\neg X_2 \vee X_4) \wedge (X_4 \vee X_5 \vee X_6)$.

Let $\mathbf{x}$ be a global strategy, then the utility functions are defined as follows. For each player $c \in P_c$, her utility function $u_c$ is such that

**(i)** $u_c(\mathbf{x}) = 3$ if $c$ plays $t$, and all of her neighbors play an action in $\{t, f\}$ in such a way that at least one of them makes the corresponding clause true;

**(ii)** $u_c(\mathbf{x}) = 2$ if $c$ plays $u$, and all of her neighbors play an action in $\{t, f\}$ in such a way that none of them makes the corresponding clause true;

**(iii)** $u_c(\mathbf{x}) = 2$ if $c$ plays $f$ and there exists $v \in Neigh(c)$ such that $v$ plays $u$;

**(iv)** $u_c(\mathbf{x}) = 1$ in all the other cases.

For each player $v \in P_v$, her utility function $u_v$ is such that

**(v)** $u_v(\mathbf{x}) = 3$ if $v$ plays an action in $\{t, f\}$ and all of her neighbors play an action in $\{t, f\}$;

**(vi)** $u_v(\mathbf{x}) = 2$ if $v$ plays $u$ and there exists $c \in Neigh(v)$ such that $c$ plays $u$;

**(vii)** $u_v(\mathbf{x}) = 1$ in all the other cases.

We claim:  $\Phi$ is satisfiable $\Leftrightarrow$ $\mathcal{G}$ admits a Nash equilibrium.

($\Rightarrow$) Assume $\Phi$ is satisfiable, and take one of such satisfying truth assignments, say $\sigma$. Consider the global strategy $\mathbf{x}$ for $\mathcal{G}$ where each player in $P_v$ chooses the action according to $\sigma$, and where each player in $P_c$ plays $t$. Note that, in this case, all players get payoff 3 according to the rules (i) and (v) above, and since 3 is the maximum payoff, $\mathbf{x}$ is a Nash equilibrium for $\mathcal{G}$.

($\Leftarrow$) Let us show that each Nash equilibrium $\mathbf{x}$ for $\mathcal{G}$ corresponds to a satisfying truth assignment of $G$. We first state the following properties of the strategies for $\mathcal{G}$.

$P_1$ : *A strategy $\mathbf{x}$ in which a player $v \in P_v$ plays $u$ cannot be a Nash equilibrium.* Indeed, assume by contradiction that $\mathbf{x}$ is a Nash equilibrium. Then, all $c \in Neigh(v)$ must play $f$ and have payoff 2, from rule (iii); otherwise they would get payoff 1, by rule (iv). However, in this case player $v$ gets payoff 1 from rule (vii), and thus, from rule (v), she can easily increase her payoff to 3 by playing an action in $\{t, f\}$. Contradiction.

$P_2$ : *A strategy $\mathbf{x}$ in which a player $c \in P_c$ plays $u$ cannot be a Nash equilibrium.* Indeed, if there is such a player $c$ that chooses $u$, then from rule (vi) each variable player $v \in Neigh(c)$ must play $u$, in order to get her maximum possible payoff 2 for the case at hand. Therefore, $\mathbf{x}$ cannot be a Nash equilibrium, by property $P_1$ above.

$P_3$ : *A strategy $\mathbf{x}$ in which all players play an action in $\{t, f\}$ and the corresponding truth assignment makes a clause $c$ false cannot be a Nash equilibrium.* In fact, in this case, from rule (ii), $c$ should play $u$ in order to get its maximum possible payoff 2, and hence $\mathbf{x}$ is not a Nash equilibrium, by property $P_2$.

$P_4$ : *A strategy $\mathbf{x}$ in which all players play an action in $\{t, f\}$ and there exists a player $c \in P_c$ that plays $f$ cannot be a Nash equilibrium.* Indeed, if $\mathbf{x}$ is a Nash equilibrium and all players play an action in $\{t, f\}$, by property $P3$ the truth assignment corresponding to this strategy satisfies each clause $c$. It follows that a player $c$ that plays $f$ contradicts the assumption that $\mathbf{x}$ is a Nash equilibrium, because $c$ could change her choice to $t$, improving her payoff to 3.

From these properties, it follows that every Nash equilibrium of $\mathcal{G}$ should be a strategy where all players play either $t$ or $f$ and all players corresponding to clauses must play $t$ and get payoff 3, as they are all made true by the truth assignment of their neighbors.

Combined with the "$\Rightarrow$"-part, this entails that there is a one-to-one correspondence between satisfying truth assignments to the variables of $\Phi$ and Nash equilibria of the game $\mathcal{G}$.

Finally, observe that the tables (matrices) representing the entries of the utility functions (rules (i)–(vii) above) can be built in polynomial time from $\Phi$. Moreover, from the assumptions that we made on the structure of $\Phi$, each player depends on 3 other players at most. $\qquad\square$

It is worthwhile noting that, though quite limited, in the above proof the interaction among players is rather complicated. In particular, it is easy to see that the dependency graph of the game described in the hardness part of the proof is cyclic. Thus, one may wonder if the problem is any easier for games with simple structured interactions. We next show that, in the general setting, even if the structure of player interactions is very simple, the problem of deciding whether pure Nash equilibria exist remains hard.

**Theorem 3.2** *Deciding whether a game $\mathcal{G}$ in general form has a pure Nash equilibrium is* NP-*complete, even if both its dependency graph and its associated hypergraph are acyclic, and the number of actions is fixed.*

**Proof**. Membership follows from the previous theorem. We next prove that this problem is NP-hard, via a reduction from SAT. Given a Boolean formula $\Phi$ over variables $X_1, ..., X_m$, we define a game $\mathcal{G}$ with $m$ players $X_1, ..., X_m$ corresponding to the variables of $\Phi$, and two additional players $T$ and $H$. Any player $X_i$, $1 \le i \le m$, has only two available actions, $t$ and $f$, corresponding to truth assignments to the corresponding variable of $\Phi$. Moreover, the utility function of each player $X_i$ is a constant, say 1. Hence, the choice of $X_i$ is independent of any other player.

The actions of player $T$ are $s$ and $u$, which can be read "satisfied" and "unsatisfied," while the actions of player $H$ are $g$ and $b$, and can be read "good" and "bad," respectively. The role of $T$ is to check whether the actions chosen by $X_1, \ldots, X_m$ encode a satisfying truth assignment for $\Phi$. Indeed, their behaviors – described below – ensure that only strategies where $T$ plays $s$ may be Nash equilibria, because of her interaction with player $H$, whose role is to discard bad strategies. Given a combined strategy $\mathbf{x}$ for the "variable players" $X_1, ..., X_m$, we denote by $\Phi(\mathbf{x})$ the evaluation of $\Phi$ on the truth values determined by the strategy $\mathbf{x}$.

**Player** $T$ depends on the players in $\{X_1, \ldots, X_m, H\}$, and her utility function is defined as follows. For any combined strategy $\mathbf{y} = \mathbf{x}_1 \cup \mathbf{x}_2$ for $X_1, \ldots, X_m, H, T$, where $\mathbf{x}_1$ is a combined strategy for $X_1, ..., X_m$ and $\mathbf{x}_2$ for $T$ and $H$:

- $u_T(\mathbf{y}) = 1$ if $\Phi(\mathbf{x}_1)$ is true and $T$ plays $s$, or $\Phi(\mathbf{x}_1)$ is false and $\mathbf{x}_2 = \{T_u, H_g\}$, or $\Phi(\mathbf{x}_1)$ is false and $\mathbf{x}_2 = \{T_s, H_b\}$;

- $u_T(\mathbf{y}) = 0$, otherwise.

**Player** $H$ depends only on $T$ and, for any combined strategy $\mathbf{x}$ for $H$ and $T$, her utility function is the following:

- $u_H(\mathbf{x}) = 1$ if $\mathbf{x}$ is either $\{T_s, H_g\}$ or $\{T_u, H_b\}$;

- $u_H(\mathbf{x}) = 0$, otherwise.

We claim there is a one-to-one correspondence between Nash equilibria of this game and satisfying assignments for $\Phi$. Indeed, let $\phi$ be a satisfying assignment for $\Phi$ and $\mathbf{x}_\phi$ the combined strategy for $X_1, \ldots, X_m$ where these players choose their actions according to $\phi$. Then, $x_\phi \cup \{T_s, H_g\}$ is a Nash equilibrium for $\mathcal{G}$, because all players get their maximum payoff.

On the other hand, if $\Phi$ is unsatisfiable, for any combined strategy $\mathbf{x}_1$ for $X_1, \ldots, X_m$, $\Phi(\mathbf{x}_1)$ is false. In this case, $T$ and $H$ have opposite interests and it is easy to check that, for each combined strategy $\mathbf{x}_2 \in St(\{H, T\})$, $\mathbf{x}_1 \cup \mathbf{x}_2$ is not a Nash equilibrium for $\mathcal{G}$, because either $H$ or $T$ can improve its payoff.

Finally, observe that the dependency graph $\mathrm{G}(\mathcal{G})$ is a tree, and that the hypergraph $\mathrm{H}(\mathcal{G})$ is acyclic. □

As shown in Figure 2, the above NP-hardness result immediately extends to all generalizations of acyclicity. The case of acyclic-hypergraph games in GNF will be dealt with in Section 4.

Let us now draw our attention to Pareto equilibria. By Definition 2.2, a Pareto Nash equilibrium exists if and only if a Nash equilibrium exists. Therefore, from Theorems 3.1 and 3.2, we get the following corollary.

**Corollary 3.3** *Deciding whether a game $\mathcal{G}$ has a Pareto Nash equilibrium is NP-complete. Hardness holds even if $\mathcal{G}$ has a fixed number of actions and if either $\mathcal{G}$ is in graphical normal form and has $k$-bounded neighborhood, for any fixed constant $k \geq 3$, or if both $\mathrm{G}(\mathcal{G})$ and $\mathrm{H}(\mathcal{G})$ are acyclic.*

However, while checking whether a global strategy $\mathbf{x}$ is a pure Nash equilibrium is tractable, it turns out that checking whether $\mathbf{x}$ is a Pareto Nash equilibrium is a computationally hard task. In fact, we next show that this problem is as difficult as checking whether $\mathbf{x}$ is a strong Nash equilibrium. However, we will see that deciding the existence of a strong Nash equilibrium is much harder, and in fact complete for the second level of the Polynomial Hierarchy. To this end, in the following proofs, we associate quantified Boolean Formulas having two quantifier alternations (2QBFs) with games.

**Quantified Boolean Formulas (QBFs) and games.** Let

$$\Xi \quad = \quad \exists \alpha_1, \ldots \alpha_n \quad \forall \beta_1, \ldots \beta_q \qquad \Phi$$

be a quantified Boolean formula in disjunctive normal form, i.e., $\Phi$ is a Boolean formula of the form $d_1 \vee \ldots \vee d_m$ over the variables $\alpha_1, \ldots \alpha_n, \beta_1, \ldots \beta_q$, where each $d_i$ is a conjunction of literals. Deciding the validity of such formulas is a well-known $\Sigma_2^P$-complete problem – (e.g. Stockmeyer & Meyer, 1973), and it is easy to see that hardness result holds even if each disjunct $d_j$ in $\Xi$ contains three literals at most and each variable occurs in three disjuncts at most. Moreover, without loss of generality, we assume that the number $m$ of disjuncts is a power of 2, say $m = 2^\ell$, for some integer $\ell \geq 2$. Note that, if $2^{\ell-1} < m < 2^\ell$, then we can build in polynomial time a new QBF $\Xi'$ having $2^l - m$ more disjuncts, each one containing both a fresh existentially quantified variable and its negation. Clearly, such disjuncts cannot be made true by any assignment, and hence $\Xi'$ is equivalent to $\Xi$. Hereafter, we will consider quantified Boolean formulas of this form, that we call R2QBF. For each such formula $\Xi$, a truth value assignment $\sigma$ to the existentially quantified variables $\alpha_1, \ldots, \alpha_n$ such that the formula $\forall \quad \beta_1, \ldots \beta_q \quad \sigma(\Phi)$ is valid is called a *witness of validity* for $\Xi$.

As a running example for this section, we consider the following QBF $\Xi_r$: $\exists \alpha_1 \alpha_2 \alpha_3 \forall \beta_1 \beta_2 \beta_3 \beta_4 \beta_5 \ (\alpha_1 \wedge \alpha_2) \vee (\alpha_1 \wedge \alpha_3) \vee (\alpha_1 \wedge \neg \beta_1) \vee (\beta_1) \vee (\neg \beta_2 \wedge \neg \beta_3) \vee (\beta_1 \wedge \beta_3) \vee (\beta_3 \wedge \beta_4) \vee (\beta_5)$.

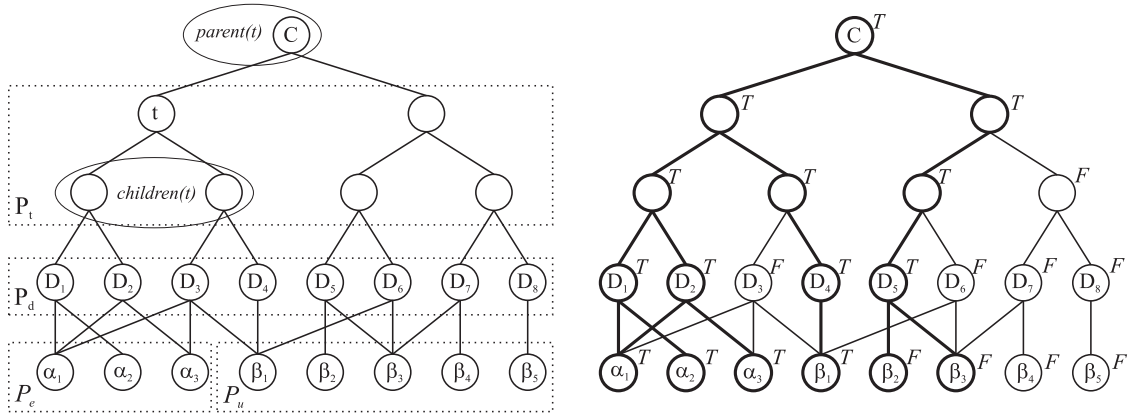

Figure 7: On the left: the dependency graph of the game $\mathcal{G}_{\Xi_r}$. On the right: a truth-value assignment for $\Xi_r$ corresponding to a strong Nash equilibrium of $\mathcal{G}_{\Xi_r}$.

We define a GNF game $\mathcal{G}_{\Xi}$ associated with a R2QBF $\Xi$ as follows. The players of $\mathcal{G}_{\Xi}$ are partitioned in five sets $P_e$, $P_u$, $P_d$, $P_t$, and the singleton $\{C\}$.

Players in $P_e$, $P_u$, and $P_d$ correspond to the existential variables $\alpha_1, \ldots \alpha_n$, to the universal variables $\beta_1, \ldots \beta_l$, and to the $m$ disjuncts of $\Phi$, respectively. Each "variable" player $v$ in $P_e \cup P_u$ may play a "truth value" action in $\{T, F\}$ (read: $\{true, false\}$), and her neighbors are the (at most three) players in $P_d$ corresponding to the disjuncts of $\Phi$ where $v$ occurs. Each "disjunct" player $p$ may play an action in $\{T, F, w\}$, and her neighbors are the (at most three) players corresponding to her variables, plus one player belonging to the set $P_t$, as described below. As shown in Figure 7, these disjunct players are the leaves of a complete binary tree comprising all players in $P_t$ as intermediate vertices, and the player $C$ as its root. In fact, the "tree" players $P_t$ act as logical-or gates of a circuit. For the sake of a simpler presentation, for each tree player $p$, $children(p)$ denotes the set of two players that are children of $p$ in this tree, while $parent(p)$ denotes her parent. As for disjunct players, the set of available actions for players in $P_t$ is $\{T, F, w\}$. Finally, the player $C$, called "challenger," may play actions in $\{T, w\}$. As shown in Figure 7, the neighbors of $C$ are the two top level tree-players.

Let $\mathbf{x}$ be a global strategy. The *utility functions* for the players in $\mathcal{G}_{\Xi}$ are defined as follows.

**Existential-variables players.** For each $\alpha \in P_e$,

    **(E-i)** $u_\alpha(\mathbf{x}) = 1$, no matter of what other players do;

**Universal-variables players.** For each $\beta \in P_u$,

    **(U-i)** $u_\beta(\mathbf{x}) = 2$, if there exists a (disjunct) neighbor playing $w$;

    **(U-ii)** $u_\beta(\mathbf{x}) = 1$ in all other cases.

**Disjuncts players.** For each $d \in P_d$,

**(D-i)** $u_d(\mathbf{x}) = 2$ if $d$ and her parent (i.e., a node from $P_t$) both play $w$, and the disjunct of $\Phi$ corresponding to $d$ is made false by the truth-value actions in $\mathbf{x}$ of her variable players, i.e., by the players in $Neigh(d) \cap (P_e \cup P_u)$;

**(D-ii)** $u_d(\mathbf{x}) = 1$ if $d$ plays $T$ and the disjunct of $\Phi$ corresponding to $d$ is made true by the truth-value actions in $\mathbf{x}$ of her variable players;

**(D-iii)** $u_d(\mathbf{x}) = 1$ if $d$ plays $F$, the disjunct of $\Phi$ corresponding to $d$ is made false by the truth-value actions of her variable players, and her parent node (a tree player) does not play $w$;

**(D-iv)** $u_d(\mathbf{x}) = 0$, in all other cases.

**Tree players.** For each $p \in P_t$,

**(TREE-i)** $u_p(\mathbf{x}) = 2$ if both $p$ and all of her neighbors play $w$;

**(TREE-ii)** $u_p(\mathbf{x}) = 1$ if $p$ plays $T$, none of her neighbors plays $w$, and some player in $children(p)$ plays $T$;

**(TREE-iii)** $u_p(\mathbf{x}) = 1$ if $p$ plays $F$, none of her neighbors plays $w$, and all players in $children(p)$ plays $F$;

**(TREE-iv)** $u_p(\mathbf{x}) = 1$ if $p$ plays an action in $\{T, F\}$, and some of her neighbors plays $w$, but not all of them;

**(TREE-v)** $u_p(\mathbf{x}) = 0$ in all other cases.

**Challenger.** For player $C$,

**(CHALL-i)** $u_C(\mathbf{x}) = 2$ if $C$ plays $w$, and either both of her neighbors play $F$, or at least one of them plays $w$;

**(CHALL-ii)** $u_C(\mathbf{x}) = 1$ if $C$ plays $T$, some of her neighbors plays $T$, and none plays $w$;

**(CHALL-iii)** $u_C(\mathbf{x}) = 0$ in all other cases.

Intuitively, universal-variable players corresponding to the variables $\beta_1, \ldots \beta_q$ choose their actions trying to falsify the formula, since their maximum payoff 2 can be obtained only if $\Phi$ is not satisfied. The strategies of variable players are suitably "evaluated" by players in $P_d \cup P_t$, and eventually by $C$.

It is worthwhile noting that $\mathcal{G}_\Xi$ can be built in polynomial time (actually, in LOGSPACE) from $\Xi$, and that each player in $\mathcal{G}_\Xi$ may play at most three actions and has a bounded number of neighbors. More precisely, for the sake of presentation, in the above construction each player has at most four neighbors. However, we will show later in this section how this bound can be reduced easily to three.

Let $\mathbf{x}$ be a global strategy for $\mathcal{G}_\Xi$. We denote by $\sigma(\mathbf{x})$ the truth-value assignment for $\Phi$ determined by the actions chosen by the variable players (i.e., those in $P_e \cup P_u$) in the strategy $\mathbf{x}$. Moreover, we denote by $\sigma_e(\mathbf{x})$ and $\sigma_u(\mathbf{x})$ its restriction to the existential and the universal variables, respectively.

**Lemma 3.4** *There is a one-to-one correspondence among the satisfying truth-value assignments for $\Phi$ and the Nash equilibria of $\mathcal{G}_\Xi$ where no player plays $w$.*

**Proof.** Assume $\Phi$ is satisfiable, and let $\sigma$ be a satisfying truth-value assignment for it. Consider the following global strategy $\mathbf{x}_\sigma$ for $\mathcal{G}_\Xi$: each variable player in $P_e \cup P_u$ chooses her truth-value action according to $\sigma$; each disjunct player in $P_d$ plays either $T$ or $F$, depending on the logical evaluation of the disjunct associated with her; each tree player $p$ in $P_t$ plays either $T$ or $F$, acting as an OR gate having as its input the values played by $children(p)$; the player $C$ plays $T$. Note that no player chooses $w$ in $\mathbf{x}_\sigma$.

Figure 7 shows on the right the strategy for $\mathcal{G}_\Xi$ associated with the truth assignment $\alpha_1 = \alpha_2 = \alpha_3 = \beta_1 = T$ and $\beta_2 = \beta_3 = \beta_4 = \beta_5 = F$.

Since $\sigma$ is a satisfying assignment, at least one disjunct of $\Phi$ will be evaluated to *true* and thus at least one of the tree-player neighbors of $C$ plays $T$. Therefore, it is easy to check that, according to the utility functions of $\mathcal{G}_\Xi$, all players get payoff 1 with respect to $\mathbf{x}_\sigma$. In particular, this follows from rule (D-ii) or (D-iii) for disjunct players, from rule (TREE-ii) or (TREE-iii) for tree players, and from rule (CHALL-ii) for the player $C$.

Moreover, the only rules that may increase some payoff from 1 to 2 are (TREE-i), (CHALL-i) and (D-i). However, no single player can increase her payoff by changing her action, because all these rules may be applied only if there is some neighbor that is playing $w$, which is not the case in $\mathbf{x}_\sigma$. It follows that the global strategy $\mathbf{x}_\sigma$ is a Nash equilibrium for $\mathcal{G}_\Xi$.

We next prove the converse, that is, we show that each Nash equilibrium $\mathbf{x}$ for $\mathcal{G}_\Xi$ where no player chooses $w$ corresponds to a satisfying truth-value assignment for $\Phi$. This proof is based on the following properties of $\mathbf{x}$:

$P_1$: *At least one player in $Neigh(C)$ does not play $F$.* Otherwise, $C$ would play $w$ in order to get payoff 2, from rule (CHALL-i), contradicting the hypothesis on $\mathbf{x}$.

$P_2$: *At least one player in $P_d$ plays $T$.* Otherwise, since no player chooses $w$ in $\mathbf{x}$, the only possible choice for all disjunct players would be $F$. However, in this case, the best available choice for all tree-players depending on disjunct players in $P_d$ is to play $F$, according to (TREE-iii). It follows by induction that all players in $P_t$ would play $F$ and, in particular, all neighbors of $C$. However, this contradicts property $P_1$ of $\mathbf{x}$.

$P_3$: *$\Phi$ is satisfied by the truth-assignment $\sigma(\mathbf{x})$.* Let $p \in P_d$ be a disjunct player that plays $T$, and whose existence is guaranteed by property $P_2$. From the hypothesis that no player chooses $w$, rule (D-i) is not applicable to $p$. It follows that the disjunction $d$ of $\Phi$ corresponding to player $p$ is true with respect to $\sigma(\mathbf{x})$. Otherwise, $\mathbf{x}$ would not be a Nash equilibrium, because $p$ would get payoff 0 from (D-iv) and could improve it by playing the correct evaluation $F$, after rule (D-iii).

Therefore, in case no player chooses $w$, all global strategies that are Nash equilibria correspond to satisfying assignments for $\Phi$. $\qquad\square$

**Lemma 3.5** *A Nash equilibrium $\mathbf{x}$ for $\mathcal{G}_\Xi$ where no player chooses $w$ is strong if and only if $\sigma_e(\mathbf{x})$ is a witness of validity for $\Xi$.*

**Proof.** ($\Rightarrow$) Assume $\mathbf{x}$ is a strong Nash equilibrium for $\mathcal{G}_\Xi$ where no player chooses $w$. Then, $\sigma(\mathbf{x})$ satisfies $\Phi$, from the previous lemma. Assume by contradiction that $\sigma_e(\mathbf{x})$ is not a witness of validity for $\Xi$. Then, there is an assignment $\sigma'_u$ for the universal variables such that $\Phi$ is not satisfied with respect to $\sigma_e(\mathbf{x}) \cup \sigma'_u$. Let $K$ be the coalition comprising all players in $\mathcal{G}_\Xi$ but the existential players in $P_e$, and let $\mathbf{y}$ be the combined strategy for $K$ such that all universal players in $P_u$ choose their truth-value action according to $\sigma'_u$, and all the other players in $K$ play $w$. By the choice of $\sigma'_u$, all the disjuncts are made false via the truth-value actions chosen by players in $P_u$. Then, from rule (D-i), all disjunct players get payoff 2 according to $\mathbf{x}_{-K}[\mathbf{y}]$. Similarly, from (Tree-i) and (Chall-i), all tree-players and the player $C$ get payoff 2 in $\mathbf{x}_{-K}[\mathbf{y}]$. However, this means that all players in the coalition are improving their payoff, and this contradicts the fact that $\mathbf{x}$ is a strong Nash equilibrium.

($\Leftarrow$) Assume that $\Xi$ is valid and let $\sigma(\mathbf{x})$ be a satisfying truth-value assignment for $\Phi$ such that $\sigma_e(\mathbf{x})$ is a witness of validity for $\Xi$, for some Nash equilibrium $\mathbf{x}$ for $\mathcal{G}_\Xi$. Indeed, from Lemma 3.4 we know such an equilibrium (where no player plays $w$) exists for every satisfying assignment, and that no player chooses $w$ in $\mathbf{x}$. Moreover, it is easy to check that all players get payoff 1, according to this global strategy. Assume by contradiction that $\mathbf{x}$ is not a strong Nash equilibrium for $\mathcal{G}_\Xi$. Then, there is a coalition $K$ and a combined strategy $\mathbf{y}$ for $K$ such that all players in the coalition may improve their payoff in the global strategy $\mathbf{x}_{-K}[\mathbf{y}]$, and hence they get payoff 2 in this strategy. Note that no existential player may belong to $K$ and thus change her action, because there is no way for her to improve her payoff. Then, from rule (Tree-i), the only way for players in $P_t$ to improve their payoff to 2 is that all of them change their actions to $w$, because all these rules require that, for each player, all of her neighbors play $w$. Thus, if one of them belongs to $K$, then all of them belong to this coalition, as well as player $C$ and all the disjunct players in $P_d$, that are their neighbors and should change her choices to $w$ in order to get 2, too. On the other hand, for all disjunct players $p$ in $P_d$, this improvement to 2 depends also on the variable players occurring in the disjunct of $\Phi$ associated with $p$ (D-i). In particular, besides playing $w$, this disjunct should also be made false, because of some change in the choices of the universal players $p$ depends on. Note that such players may in turn improve their payoff to 2, if $p$ plays $w$. It follows any coalition $K$ that shows $\mathbf{x}$ is not a Nash equilibrium contains a number of universal players that are able to let all the disjunct players to be unsatisfied and hence to change their actions to $w$, thus getting payoff 2. However, the truth-values corresponding to the actions of players in $P_u \cap K$ determine an assignment $\sigma'_u$ that contradicts the validity of $\Xi$. □

**Theorem 3.6** *Given a game $\mathcal{G}$ and a global strategy $\mathbf{x}$, deciding whether $\mathbf{x} \in \mathcal{SNE}(\mathcal{G})$ (resp., $\mathbf{x} \in \mathcal{PNE}(\mathcal{G})$) is* co-NP-*complete. Hardness holds even if the given strategy $\mathbf{x}$ is a pure Nash equilibrium, $\mathcal{G}$ is in graphical normal form and has 3-bounded neighborhood, and the number of actions is fixed.*

**Proof.** *Membership.* Deciding whether $\mathbf{x} \notin \mathcal{PNE}(\mathcal{G})$ is in NP: (i) check in polynomial time whether $\mathbf{x} \notin \mathcal{NE}(\mathcal{G})$; (ii) if this is not the case, guess a global strategy $\mathbf{y}$ and check in polynomial time whether $\mathbf{y} \in \mathcal{NE}(\mathcal{G})$ and $\mathbf{y}$ dominates $\mathbf{x}$. Similarly, deciding whether $\mathbf{x} \notin \mathcal{SNE}(\mathcal{G})$ is in NP: (i) check in polynomial time whether $\mathbf{x} \notin \mathcal{NE}(\mathcal{G})$; (ii) if this is not the case, guess a coalition of players $K$ and a combined strategy $\mathbf{y}$ for the players in $K$, and check in polynomial time whether all the players in $K$ increase their payoff by playing their actions according to the new strategy $\mathbf{y}$.

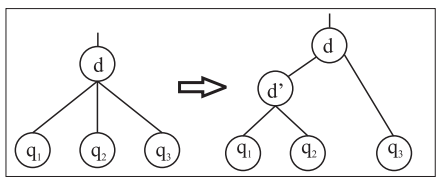

Figure 8: Transformation of a disjunct containing exactly three literals.

*Hardness.* It is well-known that checking whether a satisfiable formula $\Phi$ in 3DNF is *tautologically valid* is co-NP complete. We reduce this problem to the problem of checking whether a given Nash equilibrium is strong (Pareto). Let $\Phi$ be a satisfiable formula in 3DNF. Find a satisfying truth value assignment $\sigma$ for $\Phi$. This is obviously possible in polynomial time, given that $\Phi$ is satisfiable and in DNF. Define $\Xi = \forall \beta_1, \dots \beta_q \quad \Phi$. $\Xi$ can be considered a (degenerate) R2QBF without existentially quantified variables. Let $\mathcal{G}_\Xi$ be the game associated with $\Xi$. Obviously, $\mathcal{G}_\Xi$ can be constructed in polynomial time from $\Phi$. Let $\mathbf{x}$ be the Nash equilibrium for $\mathcal{G}_\Xi$ determined by this truth-value assignment where no player chooses $w$, as described in Lemma 3.4, and note that even this equilibrium can be computed in polynomial time from $\sigma$. Then, by Lemma 3.5, $\mathbf{x}$ is strong if only if $\Phi$ is valid (i.e., the vacuous assignment $\sigma_e$ is a witness of validity). This settles the hardness of checking whether a given Nash Equilibrium is strong. In addition, it is not hard to see that in the above construction $\mathbf{x}$ is a Pareto equilibrium for $\mathcal{G}_\Xi$ if and only if $\Phi$ is a tautology. Indeed, note that, there is a truth-value assignment $\sigma'$ that falsifies $\Phi$, if and only if there is a global strategy $\mathbf{y}$ where (universal) variable players play according to $\sigma'$ and all other players choose $w$, such that $\mathbf{y}$ dominates the Nash equilibrium $\mathbf{x}$. Thus checking whether a given Nash equilibrium is Pareto is co-NP complete, too.

We conclude the proof by observing that, in our reduction, each player in $P_e \cup P_u \cup P_t \cup \{C\}$ depends on three other players at most, but some player $d$ in $P_d$ may depend on four other players, if the corresponding disjunct contains exactly three literals. In this case, we may introduce an additional player $d'$ whose set of actions is $\{T, F, w\}$ and whose neighbors are the first two literals plus $d$. Moreover, her utility function is such that $d'$ acts as an AND-gate on the inputs of the literals, preferring $w$ if the two literals are evaluated false, or if $d$ plays $w$. In this new encoding, $d$ depends only on $d'$ and on the third literal occurring in the disjunct. As for the basic construction previously defined, her payoff is 2 if she plays $w$, her parent plays $w$, and the third literal is false, or if $d'$ plays $w$. Figure 8 shows this transformation. Note that this construction preserves all the properties proved so far, and each player depends on three other players at most. $\qquad\square$

Deciding whether a game has a strong Nash equilibrium turns out to be much more difficult than deciding the existence of pure (Pareto) Nash equilibria. Indeed, this problem is located at the second level of the polynomial hierarchy.

**Theorem 3.7** *Given a game $\mathcal{G}$, deciding whether $\mathcal{G}$ has a strong Nash equilibrium is $\Sigma_2^P$-complete. Hardness holds even if $\mathcal{G}$ is in graphical normal form and has 3-bounded neighborhood, and the number of actions is fixed.*

**Proof.** *Membership.* We can decide that $\mathcal{SNE}(\mathcal{G}) \neq \emptyset$ by guessing a global strategy $\mathbf{x}$ and verifying that $\mathbf{x}$ is a strong Nash equilibrium for $\mathcal{G}$. From Theorem 3.6, the latter task is feasible in co-NP. It follows that the problem belongs to $\Sigma_2^P$.

*Hardness.* Let $\Xi = \exists \alpha_1, \ldots \alpha_n \forall \beta_1, \ldots \beta_q \Phi$ be a R2QBF. Recall that deciding the validity of such formulas is a $\Sigma_2^P$-complete problem – see the definition of R2QBF above, in the current section.

We define a game $\mathcal{G}'_\Xi$ associated with $\Xi$, obtained from $\mathcal{G}_\Xi$ with the addition of one more gadget. In $\mathcal{G}'_\Xi$ there is a fresh player $D$ depending on the player $C$ only. It is called "duplicator," because $D$ gets her maximum payoff if she plays the same action as the challenger player $C$. On the other hand, also $C$ depends on this new player $D$, besides her other tree-players neighbors (recall the construction shown in Figure 7). Both $C$ and $D$ play actions in $\{T, w, u\}$. Everything else is the same as in $\mathcal{G}_\Xi$, but for the utility functions of players $C$ and $D$:

**Challenger.** For player $C$,

> **(CHALL-i)** $u_C(\mathbf{x}) = 2$ if $C$ and $D$ play different actions in $\{w, u\}$ and either all of her tree-players neighbors (i.e., the players in $Neigh(C) - \{D\}$) play $F$, or at least one of them plays $w$;

> **(CHALL-ii)** $u_C(\mathbf{x}) = 1$ if $C$ plays $T$, one player in $Neigh(C) - \{D\}$ plays $T$, and none plays $w$;

> **(CHALL-iii)** $u_C(\mathbf{x}) = 0$ in all other cases.

**Duplicator.** For player $D$,

> **(DUPL-i)** $u_D(\mathbf{x}) = 1$ if $D$ plays the same action as $C$;

> **(DUPL-ii)** $u_D(\mathbf{x}) = 0$ in all other cases.

Recall that, in order to maximize their payoffs, the universal-variable players try to make $\Phi$ false, and that the strategies of variable players are suitably "evaluated" by players in $P_d \cup P_t$, acting as a Boolean circuit. Moreover, the challenger and the duplicator are designed in such a way that strategies where some player chooses $w$, which do not correspond to satisfying assignments for $\Phi$, cannot lead to Nash equilibria. Formally, for any Nash equilibrium of $\mathcal{G}'_\Xi$, the following properties hold.

$P_1$ : *At least one player in $Neigh(C) - \{D\}$ does not play $F$ and no player in $Neigh(C) - \{D\}$ plays $w$.* Otherwise, $C$ would play an action in $\{w, u\}$ different from the one played by $D$ in $\mathbf{x}$, in order to get payoff 2, from rule (CHALL-i). However, such a strategy cannot be an equilibrium, because $D$ could improve her payoff by choosing the same action as $C$ in $\mathbf{x}$.

$P_2$ : *If a player in $P_d \cup P_t$ plays $w$ then all players in $P_d \cup P_t$ play $w$.* Indeed, let $p$ be a player in $P_d \cup P_t$ playing $w$ and assume by contradiction that there is a player $q$ in $P_d \cup P_t$ that does not play $w$. W.l.o.g., we can assume that $q \in Neigh(p)$ and viceversa. Then, $p$ gets payoff 0, but could increase her payoff by playing an action in $\{T, F\}$ according to (D-ii) or (D-iii) for players in $P_d$, and to (TREE-ii), (TREE-iii) or (TREE-iv) for players in $P_t$. This contradicts the fact that $\mathbf{x}$ is a Nash equilibrium.

$P_3:$ *No player in $P_d \cup P_t$ plays $w$.* Otherwise, from $P_2$, all players in $P_d \cup P_t$ play $w$ and, in particular, the players in $Neigh(C) - \{D\}$. However, this is not possible, after property $P_1$.

Therefore, rule (Chall-i) is not applicable for $C$ with respect to **x**. However, from the above properties at least one of her tree-players neighbors should play $T$ and $C$ can play $T$ in her turn and get payoff 1, from rule (Chall-ii). Then, $D$ may play $T$ and get payoff 1, as well. Thus, it is no longer possible to have a Nash equilibrium with some player choosing the action $w$. Therefore, after Lemma 3.5 (as the presence of the duplicator does not affect that proof), we get the following fundamental result: a global strategy **x** is a strong Nash equilibrium for $\mathcal{G}'_\Xi$ if and only if the truth-value assignment $\sigma_e(\mathbf{x})$ is a witness of validity for $\Xi$. In particular, $\mathcal{SNE}(\mathcal{G}'_\Xi) \neq \emptyset$ if and only if $\Xi$ is valid.

Finally, observe that even the game $\mathcal{G}'_\Xi$ can be modified as shown in Figure 8, in order to get a 3-bounded neighborhood game. □

We next show that, if the game is given in general form, the above hardness result holds even if the structure of player interactions is very simple.

**Theorem 3.8** *Deciding whether a game $\mathcal{G}$ in general form has a strong Nash equilibrium is $\Sigma_2^P$-complete, even if both its dependency graph and its associated hypergraph are acyclic, and the number of actions is fixed.*

**Proof.** The proof of membership in $\Sigma_2^P$ is the same as the membership proof in the previous theorem. We next prove that this problem is hard for $\Sigma_2^P$.

Let $\Xi = \exists \alpha_1, \ldots \alpha_n \forall \beta_1, \ldots \beta_q \; \Phi$ be a quantified Boolean formula in disjunctive normal form. We define a game $\bar{\mathcal{G}}_\Xi$ with $m$ players $\alpha_1, \ldots \alpha_n, \beta_1, \ldots \beta_q$ corresponding to the existentially and universally quantified variables of $\Xi$, and two additional players $T$ and $H$. The game is based on a combination of the game techniques exploited in the proofs of Theorem 3.2 and Theorem 3.7. Any player associated with a variable has only two available actions, $t$ and $f$, which represent truth assignments to the corresponding variable of $\Phi$; the actions of player $T$ are $s$ and $u$, which can be read "satisfied" and "unsatisfied," while the actions of player $H$ are $g$ and $b$, and can be read "good" and "bad," respectively.

Given a combined strategy **x**, we denote by $\Phi(\mathbf{x})$ the evaluation of $\Phi$ on the truth values determined by the strategy **x**. Then, the utility functions are defined as follows. Any player $\alpha_i$ $(1 \leq i \leq n)$ gets always payoff 1. Any player $\beta_j$ $(1 \leq j \leq q)$ depends on $T$ and gets payoff 1 if $T$ plays $s$ in **x**, and payoff 2 if $T$ plays $u$ in **x**. Player $H$ depends only on $T$ and gets payoff 1 if **x** contains either $\{T_s, H_g\}$ or $\{T_u, H_b\}$, and 0, otherwise. Finally, player $T$ depends on players in $\{\alpha_1, \ldots \alpha_n, \beta_1, \ldots \beta_q, H\}$, and her utility function is defined as follows:

- $u_T(\mathbf{x}) = 2$, if $\Phi(\mathbf{x})$ is false, $T$ plays $u$, and $H$ plays $g$;

- $u_T(\mathbf{x}) = 1$, if $\Phi(\mathbf{x})$ is true and $T$ plays $s$, or if $\Phi(\mathbf{x})$ is false, $T$ plays $s$, and $H$ plays $b$;

- $u_T(\mathbf{x}) = 0$, otherwise.

First, observe that both the dependency graph and the dependency hypergraph of $\bar{\mathcal{G}}_\Xi$ are acyclic. Moreover, after the proof of Theorem 3.2, it is easy to see that there is a one-to-one correspondence between Nash equilibria of this game and satisfying assignments for $\Phi$. Then, for any Nash equilibrium $\mathbf{x}$ for $\bar{\mathcal{G}}_\Xi$, we denote by $\sigma^{\mathbf{x}}$ its corresponding truth-value assignment, and by $\sigma_e^{\mathbf{x}}$ the restriction of this assignment to the existential variables of $\Xi$. Note that, as shown in the above mentioned proof, at any Nash equilibrium, player $T$ plays $s$ and all players in the game get payoff 1.

We next prove that any witness of validity for $\Xi$ corresponds to a strong Nash equilibrium for $\bar{\mathcal{G}}_\Xi$. Let $\mathbf{x}$ be a Nash equilibrium and consider a coalition $K$ of players that deviate from $\mathbf{x}$, leading to a new profile $\mathbf{x}'$. From the definition of the game, the only way for the coalition to get a payoff higher than 1 is that $T$ changes her choice to $u$. In this case, if $\Phi(\mathbf{x}')$ is false (and $H$ plays $g$), $T$ will get payoff 2. Since $\Phi$ is true with respect to $\mathbf{x}$, it follows that some variable players have to change their choices, which means that they should belong to the coalition and improve their payoffs. Therefore, all variable players in $K$ should correspond to universally quantified variables, as only these players are able to improve their payoffs from 1 to 2. Thus, such a coalition $K$ exists if and only if the universally quantified variables can make the formula false, that is, $\sigma_e^{\mathbf{x}}$ is not a witness of validity for $\Xi$. Equivalently, it follows that $\mathbf{x}$ is a strong Nash equilibrium if and only if $\Xi$ is valid. □

**Remark 3.9** *Recall that we assumed any general game $\mathcal{G}$ to be taken from the class $\mathcal{C}_{\bar{k}}$.*

*It is worthwhile noting that, for games without restriction on player interactions, our hardness results hold for games where the utility functions are computable in constant time, too. Namely, consider Theorems 3.1, 3.6, and 3.7. In these constructions, each player has at most three neighbors and a fixed number of actions. Therefore, the utility function of each player is computable in constant time.*

## 4. Easy Games

Before we deal with tractable games in graphical normal form (GNF), let us recall that all computational problems dealt with in this paper are tractable for games in standard normal form (SNF), even for arbitrary interactions among players. Actually, we next point out that they can be carried out in logarithmic space and are thus in a very low complexity class that contains highly parallelizable problems only. This is not very surprising, because in fact the size of SNF games may be exponentially larger than the size of the same games encoded in GNF.

**Theorem 4.1** *Given a game in standard normal form, the following tasks are all feasible in logarithmic space: Determining the existence of a pure Nash equilibrium, a pure Pareto equilibrium, or a strong Nash equilibrium, and computing all such equilibria.*

**Proof**. Let $P$, as usual, denote the set of players. We assume w.l.o.g. that each player has at least two possible actions (in fact, a player with a single action can be eliminated from the game by a simple logspace transformation, yielding an equivalent game). The size of the input matrix is thus at least $2^{|P|}$.

Given that *all* possible global strategies are explicitly represented, each corresponding to a table cell (which can be indexed in logarithmic space in the size of the input, which corresponds to polynomial space in $|P| \times |A|$, where $P$ and $A$ are the sets of players and the set of all possible actions, respectively), the Nash equilibria are easily identified by scanning (i.e., enumerating) all global strategies $\mathbf{x}$ keeping a logspace index of $\mathbf{x}$ on the worktape, and checking in logarithmic space (in the size of the input) whether no player can improve her utility by choosing another action. Given that all Nash Equilibria can be generated in logarithmic space, they also can be generated in polynomial time.

The Pareto equilibria can be identified by successively enumerating all Nash equilibria $\mathbf{x}$, and by an additional loop for each $\mathbf{x}$, enumerating all Nash equilibria $\mathbf{y}$ (indexed in logarithmic space as above) and outputting $\mathbf{x}$ if there is no $\mathbf{y}$ such that, $\forall p \in P$, $u_p(\mathbf{y}) > u_p(\mathbf{x})$. The latter condition can be tested by means of a simple scan of the players.

Strong Nash equilibria can be identified by enumerating all Nash equilibria and by scanning all possible coalitions of the players (which can be indexed in logarithmic space, as their number is $2^{|P|}$) in order to discard those equilibria $\mathbf{x}$ for which there exists a coalition $K \subseteq P$ and a combined strategy $y$ for $K$, such that for each $p \in K$, $u_p(\mathbf{x}) < u_p(\mathbf{x}_{-K}[y])$. Finally, note that, for a fixed coalition $K$, the enumeration of all the combined strategies $y$ for $K$ can be carried out by means of an additional nested loop, requiring logarithmic space for indexing each such strategy. □

## 4.1 Constraint Satisfaction Problems and Games in Graphical Normal Form

Let us now consider games in GNF. We first establish an interesting connection between constraint satisfaction problems and games. An instance of a *constraint satisfaction problem (CSP)* (also *constraint network*) is a triple $I = (Var, U, \mathcal{C})$, where $Var$ is a finite set of variables, $U$ is a finite domain of values, and $\mathcal{C} = \{C_1, C_2, \ldots, C_q\}$ is a finite set of constraints. Each constraint $C_i$ is a pair $(S_i, r_i)$, where $S_i$ is a list of variables of length $m_i$ called the *constraint scope*, and $r_i$ is an $m_i$-ary relation over $U$, called the *constraint relation*. (The tuples of $r_i$ indicate the allowed combinations of simultaneous values for the variables $S_i$.) A *solution* of a CSP instance is a substitution $\theta : Var \longrightarrow U$, such that for each $1 \leq i \leq q$, $S_i\theta \in r_i$. The problem of deciding whether a CSP instance has any solution is called *constraint satisfiability (CS)*. Since we are interested in CSPs associated with games, where variables are players of games, we will use interchangeably the terms variable and player, whenever no confusion arises.

Let $\mathcal{G} = \langle P, Neigh, A, U \rangle$ be a game and $p \in P$ a player. Define the *Nash constraint* $NC(p) = (S_p, r_p)$ as follows: The scope $S_p$ consists of the players in $\{p\} \cup Neigh(p)$, and the relation $r_p$ contains precisely all combined strategies $\mathbf{x}$ for $\{p\} \cup Neigh(p)$ such that there is no $y_p \in St(p)$ such that $u_p(\mathbf{x}) < u_p(\mathbf{x}_{-p}[y_p])$. Thus note that, for each Nash equilibrium $\mathbf{x}$ of $\mathcal{G}$, $\mathbf{x} \cap St(S_p)$ is in $r_p$.

The constraint satisfaction problem associated with $\mathcal{G}$, denoted by $CSP(\mathcal{G})$, is the triple $(Var, U, \mathcal{C})$, where $Var = P$, the domain $U$ contains all the possible actions of all players, and $\mathcal{C} = \{NC(p) \mid p \in P\}$, i.e., it is the set of Nash constraints for the players in $\mathcal{G}$.

**Example 4.2** The constraint satisfaction problem associated with FRIENDS game is $(\{F, G, R, P, M\}, \{m, o\}, \mathcal{C})$, where the set of constraints contains exactly the following Nash constraints: $NC(F) = (\{F, P, R\}, r_F)$, $NC(G) = (\{G, P, F\}, r_G)$, $NC(R) = (\{R, F\}, r_R)$, $NC(P) = (\{P, F\}, r_P)$, and $NC(M) = (\{M, R\}, r_M)$, where the constraint scopes are shown in Figure 9. □

Figure 9: Constraint relations of the game FRIENDS in Example 2.1.

The structure of a constraint satisfaction problem $I = (Var, U, \mathcal{C})$ is represented by the hypergraph $\mathrm{H}(I) = (V, H)$, where $V = Var$ and $H = \{var(S) \mid C = (S, r) \in \mathcal{C}\}$, and $var(S)$ denotes the set of variables in the scope $S$ of the constraint $C$. Therefore, by definition of $CSP(\mathcal{G})$, the hypergraph of any game $\mathcal{G}$ coincides with the hypergraph of its associated constraint satisfaction problem, and thus they have the same structural properties.

The following theorem establishes a fundamental relationship between games and CSPs.

**Theorem 4.3** *A strategy* $\mathbf{x}$ *is a pure Nash equilibrium for a game* $\mathcal{G}$ *if and only if it is a solution of* $CSP(\mathcal{G})$.

**Proof**. Let $\mathbf{x}$ be a Nash equilibrium for $\mathcal{G}$ and let $p$ be any player. Then, for each strategy $p_a \in St(p)$, $u_p(\mathbf{x}) \geq u_p(\mathbf{x}_{-p}[p_a])$. Since $u_p$ depends only on the players in $\{p\} \cup Neigh(p)$, their combined strategy $\mathbf{x}' \subseteq \mathbf{x}$ is a tuple of $NC(p)$, by construction. It follows that the substitution assigning to each player $p$ its individual strategy $p_a \in \mathbf{x}$ is a solution of $CSP(\mathcal{G})$.

On the other hand, consider any solution $\theta$ of $CSP(\mathcal{G})$, and let $p$ be any player. Let $P' = \{p\} \cup Neigh(p)$ and $\mathbf{x}'$ the combined strategy $\{\theta(q) \mid q \in P'\}$. Then, $\mathbf{x}'$ is a tuple of $NC(p)$, because $\theta$ is a solution of $CSP(\mathcal{G})$. Thus, for each $p$, by definition of $NC(p)$, there is no individual strategy for $p$ that can increase her utility, given the strategies of the other players. It follows that the global strategy containing $\theta(p)$ for each player $p$ is a Nash equilibrium for $\mathcal{G}$. $\qquad\square$

The following theorem states the feasibility of the computation of $CSP(\mathcal{G})$.

**Theorem 4.4** *Let* $\mathcal{G}$ *be a game having small neighborhood or in graphical normal form. Then, computing* $CSP(\mathcal{G})$ *is feasible in polynomial time.*

**Proof**. Let $\mathcal{G} = \langle P, Neigh, A, U \rangle$ be a game having small neighborhood. We show that $NC(p) = (S_p, r_p)$ can be computed in polynomial time. We initialize $r_p$ with all the combined strategies for $\{p\} \cup Neigh(p)$. The number of these combined strategies is bounded by

$$maxAct(\mathcal{G})^{|Neigh(p)|} = 2^{\log(maxAct(\mathcal{G})^{|Neigh(p)|})} \leq 2^{i(\mathcal{G}) \times \log(\|\mathcal{G}\|)} = \|\mathcal{G}\|^{i(\mathcal{G})},$$

where the intricacy $i(\mathcal{G})$ of $\mathcal{G}$ is given by

$$\frac{maxNeigh(\mathcal{G}) \times \log maxAct(\mathcal{G})}{\log \|\mathcal{G}\|}.$$

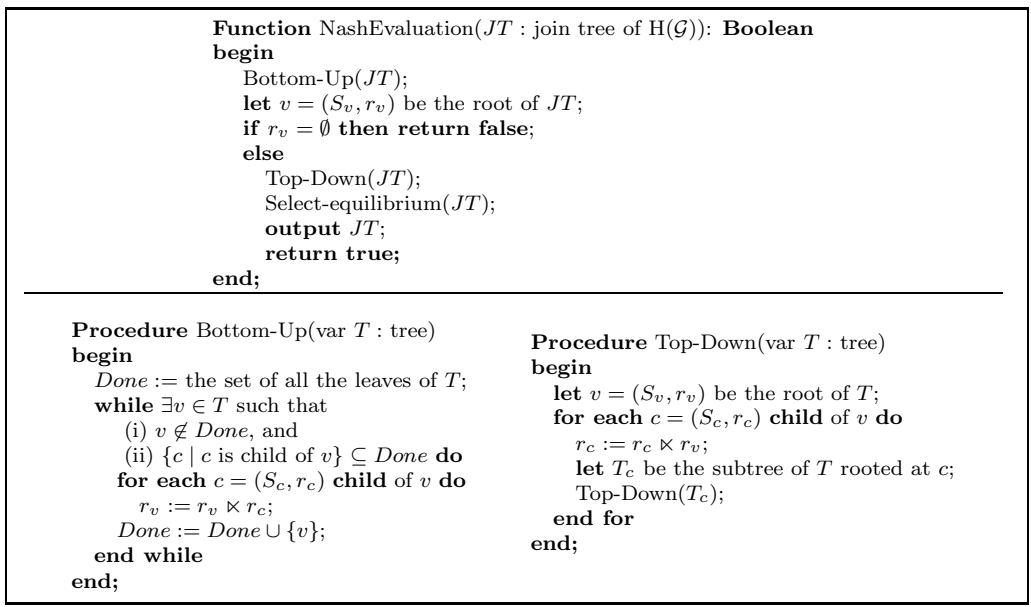

Figure 10: Evaluation of an acyclic game.

Since $\mathcal{G}$ has small neighborhood, $i(\mathcal{G})$ is bounded by some constant, and thus the set of all combined strategies for $p$ is polynomially bounded. The initialization process computes for each $p$ all corresponding combined strategies (via a simple enumeration), and thus takes polynomial time (in the size of $\mathcal{G}$).

Now, for each tuple $\mathbf{x}$ in $r_p$ we have to check whether it should be kept in $r_p$ or not. Let $m = u_p(\mathbf{x})$. For each action $a \in Act(p)$, compute in polynomial time $m' = u_p(\mathbf{x}_{-p}[p_a])$ and delete $\mathbf{x}$ if $m' > m$. It follows that $CSP(\mathcal{G})$ can be computed in polynomial time from $\mathcal{G}$.

A similar line of reasoning applies if $\mathcal{G}$ is in GNF. In this case, the utility functions are explicitly given in input in a tabular form, and thus the computation of Nash constraints is yet easier. In fact, this task is feasible in logspace for GNF games. ☐

After Theorem 4.3, an acyclic-hypergraph game $\mathcal{G}$ having small neighborhood or in graphical normal form can be solved in polynomial time. Indeed, $\mathcal{G}$ and $CSP(\mathcal{G})$ have the same hypergraph and, as shown by Gottlob et al. (2000), a solution of an acyclic constraint satisfaction problem can be computed by (a slight adaptation of) the well known Yannakakis's algorithm for evaluating acyclic conjunctive queries (Yannakakis, 1981), or by the LOGCFL algorithm proposed by Gottlob et al. (2000), which shows this problem is highly parallelizable — see Section 6, for more information on the complexity class LOGCFL. For the sake of completeness, in Figure 10, we report an algorithm for deciding the existence of a Nash equilibrium and for computing Nash equilibria of acyclic-hypergraph games, based on these results. We assume the reader is familiar with typical database operations like semi-joins (for more details, see, e.g., Maier, 1986).

The algorithm takes in input a join tree $JT$ of $\mathrm{H}(G)$. With a small abuse of notation, each vertex of $JT$, which is formally a hyperedge $H_v$ associated with a player $v$, is also used to denote the player herself as well as the Nash constraint $(S_v, r_v)$ associated with $v$.

```
Procedure Select-equilibrium(var T : tree)
begin
    let v = (S_v, r_v) be the root of T;
    select any combined strategy t_v ∈ r_v s.t. ∀t'_v ∈ r_v, u_v(t'_v) ≤ u_v(t_v);
    delete all tuples in r_v, but t_v;
    for each c = (S_c, r_c) child of v do
        r_c := r_c ⋉ r_v;
        let T_c be the subtree of T rooted at c;
        Select-equilibrium(T_c);
    end for
end;
```

Figure 11: Selection of a (Pareto) Nash equilibrium of an acyclic game.

The algorithm consists of two phases. In the first bottom-up phase, the constraint relation $r_v$ of each node $v = (S_v, r_v)$ of $JT$ is filtered by means of a semijoin with the constraint relation $r_c$ (denoted by $r_v \ltimes r_c$) of each of her children $c$ in $JT$. This semijoin eliminates all tuples from $r_v$ corresponding to combined strategies of the players in $P' = (Neigh(c) \cup \{c\}) \cap (Neigh(v) \cup \{v\}))$ that are not available (or no longer available) in $r_c$.

This way, all tuples corresponding to strategies that do not match and hence cannot lead to Nash equilibria are deleted, starting from the leaves. Finally, either the root is empty and hence $\mathcal{G}$ does not have equilibria, or the tuples remaining in the root $p$ encode the strategies that $p$ may choose in Nash equilibria of the game. In the top-down phase, this property of the root is propagated down the tree by taking the semi-join of every vertex with all its children. At the end, we get a tree such that all tuples encode strategies belonging to Nash equilibria and, vice versa, all Nash equilibria are made from strategies in the relations stored in $JT$. Then, by standard techniques developed for acyclic database queries and CSPs, we can compute from $JT$ all Nash equilibria of $\mathcal{G}$ in a backtrack-free way, and thus in time polynomial in the combined size of the input game and of the equilibria in output. Note that this is the best we can do, as a game may have an exponential number of equilibria.

For completeness, Figure 11 shows Procedure *Select-equilibrium*, that selects from $JT$ one Nash equilibrium. It is very similar to Procedure *Top-Down*, but for the selection step before semi-joins: for any vertex, *Select-equilibrium* first picks a combined strategy $t_v \in r_v$, deletes all other tuples in $r_v$, and then performs the semi-joins with its children and calls itself recursively, to propagate the choice of $t_v$ towards the leaves of the tree. Note that the selection of $t_v$ may be arbitrary, after the previous bottom-up and top-down steps. However, in Figure 11 we select strategies giving the best payoffs, in order to get a Nash equilibrium that cannot be dominated by any other Nash equilibrium.

**Theorem 4.5** *Deciding the existence of pure Nash equilibria, as well as computing a Nash equilibrium is feasible in polynomial time for all classes $\mathcal{C}$ of acyclic-hypergraph games such that every game $\mathcal{G} \in \mathcal{C}$ has small neighborhood or is in graphical normal form.*

From the above discussion, it immediately follows that this tractability result can be extended to the problem of computing a Pareto Nash equilibrium.

**Theorem 4.6** *Deciding the existence of Pareto Nash equilibria, as well as computing a Pareto Nash equilibrium for pure strategies is feasible in polynomial time for all classes $\mathcal{C}$*

```
Function InCoalition_{x,JT}(H_p : vertex, st: combined strategy): boolean
begin
    let (S_p, r_p) be the constraint associated with player p and N^+ = {p} ∪ Neigh(p);
    guess a tuple st' ∈ r_p such that st' matches st on the players they have in common;
    if p's strategy in st' is different from her strategy in x  and  u_p(x_{-N^+}[st']) ≤ u_p(x) then
        return false;
    else
        let K_p be the set of children of H_p in the join tree JT;
        if K_p = ∅ then
            return true;
        else
            return ⋀_{H_{p'} ∈ K_p} InCoalition_{x,JT}(H_{p'}, st');
    end if
end.
```

Figure 12: Algorithm for deciding the existence of a coalition improving **x**.

*of acyclic-hypergraph games such that every game $\mathcal{G} \in \mathcal{C}$ has small neighborhood or is in graphical normal form.*

**Proof.** Recall that Procedure *Select-equilibrium* in Figure 11, at each vertex $v$ encountered during the visit of $JT$, select a combined strategy that guarantees to the player corresponding to $v$ the maximum payoff over the available choices (that, at this point, are all and only those strategies that may lead to Nash equilibria). In particular, the payoff of the first player to be evaluated, say the root $p$, cannot be worse than the payoff of $p$ in any other available strategy. Thus, the tuples left in $JT$ after this procedure encode a Pareto Nash equilibrium of $\mathcal{G}$, as it cannot be strictly dominated by any other Nash equilibrium.

For the sake of completeness, we point out that, differently from the previous case of plain Nash equilibria, from $JT$ we cannot compute easily *all* Pareto Nash equilibria of the game in input-output polynomial time. □

One may thus wonder whether the above result holds for strong Nash equilibria, too. Unfortunately, we next show that computing a Strong Nash equilibrium is a difficult problem even in the case of acyclic interactions among players. However, the complexity is reduced by one level with respect to the arbitrary interaction case, because checking whether a given equilibrium is strong is feasible in polynomial time in the acyclic case.

**Lemma 4.7** *Let $\mathcal{G}$ be an acyclic-hypergraph game that has small neighborhood or is in graphical normal form, and let $\mathbf{x}$ be a global strategy. Then, deciding whether $\mathbf{x} \in \mathcal{SNE}(\mathcal{G})$ is feasible in polynomial time.*

**Proof.** Since $\mathcal{G}$ has small neighborhood or is in graphical normal form, from Theorem 4.4 we can build in polynomial time the constraints associated with each player. Moreover, its hypergraph $H(\mathcal{G})$ is acyclic and thus it has a join tree. Let $JT$ be a join tree of $H(\mathcal{G})$. We show how to use $JT$ for deciding in polynomial time whether the strategy $\mathbf{x}$ is not in $\mathcal{SNE}(\mathcal{G})$, i.e., that there exists a coalition $C$ of players getting an incentive to deviate all together from $\mathbf{x}$. Then, the result follows because PTIME is closed under complementation. Specifically, we next show an implementation of this task by an alternating Turing machine $M$ with a logarithmic-space working tape. Therefore, the problem is in ALOGSPACE, which is equal to PTIME (Chandra, Kozen, & Stockmeyer, 1981).

The machine $M$ for deciding whether $\mathbf{x}$ is not a strong Nash equilibrium works as follows:

- guess a player $q$;

- guess a strategy $st_q$ for $q$ that is different from her choice in $\mathbf{x}$;

- root the tree $JT$ at the vertex corresponding to the characteristic edge $H_q$ of player $q$;

- check that $InCoalition_{\mathbf{x},JT}(H_q, st_q)$ returns true, where $InCoalition_{\mathbf{x},JT}$ is the Boolean function shown in Figure 12.

Intuitively, the non-deterministic Turing machine first chooses a player $q$ belonging to a possible coalition $C$ disproving $\mathbf{x}$. Thus, $q$ should improve her payoff, and in general – unless $\mathbf{x}$ is not a strong Nash equilibrium, getting this improvement may require that some of her neighbors $K_q$ deviate from $\mathbf{x}$ and hence belong to $C$. However, in this case, all players in $K_q$ should be able to improve their payoffs. Again, to do that, they can involve other players in the coalition, and so on.

Whether or not this process is successful is checked by the recursive Boolean function $InCoalition_{\mathbf{x},JT}$, which takes in input a vertex $H_p$ of the join tree $JT$ and a combined strategy $st$. Recall that each vertex of the join tree is an (hyper)edge of the hypergraph, corresponding to a player of $\mathcal{G}$. In particular, $H_p$ is the characteristic edge of player $p$. $InCoalition_{\mathbf{x},JT}$ has to check whether all players deviating from the given global strategy $\mathbf{x}$ are able to improve their payoffs. At the first call of this function, the first parameter $H_q$ is the root of $JT$ and identifies the first player $q$ chosen for the coalition $C$. The second parameter $st$ is the strategy chosen by $q$, which is different from $q$'s corresponding choice in $\mathbf{x}$. At a generic recursive call, the first parameter $H_p$ identifies a player $p$ to be checked, and the second parameter $st$ encodes a combined strategy for the player $w$ associated with the parent $H_w$ of $H_p$ in $JT$ and for $w$'s neighbors. Now, the function has to check that either $p$ does not change her choice with respect to $\mathbf{x}$, or she changes and improves her payoff. To this end, the function guesses a tuple $st' \in r_p$, where $r_p$ is the constraint relation associated with $p$. Then, $st'$ encodes a combined strategy for $p$ and her neighbors. This strategy has to match the parameter $st$ on the players they have in common, because $st$ contains the actions already chosen by the algorithm when the parent $H_w$ of $H_p$ has been evaluated. Then, if $p$'s choice in $st'$ is different from $p$'s choice in $\mathbf{x}$, it means that $p$ has been non-deterministically chosen as a member of the coalition $C$. Thus, $p$ should improve her payoff, or she immediately causes the fail of this computation branch of the nondeterministic Turing machine. Otherwise, that is, if $p$ plays the same action as in $\mathbf{x}$, then she does not belong to $C$ and the function has to check, recursively, that in the rest of the join tree all deviating players improve their payoffs. This is done by propagating the current combined strategy $st'$ to the children of $H_p$ in $JT$. Observe that this propagation is necessary even if $p$ does not belong to $C$, because connected coalitions do not necessarily induce connected subtrees in $JT$. Indeed, it may happen that some player $z$ belonging to the coalition is a neighbor of both $p$ and $w$, but her characteristic edge $H_z$ occurs far from $H_w$ in the join tree, possibly in the subtree of $JT$ rooted at $p$. (For the sake of completeness, note that in this case $z$ should be a neighbor of all players occurring in the path from $H_z$ to $H_w$ in $JT$, from the connectedness property of join trees.)

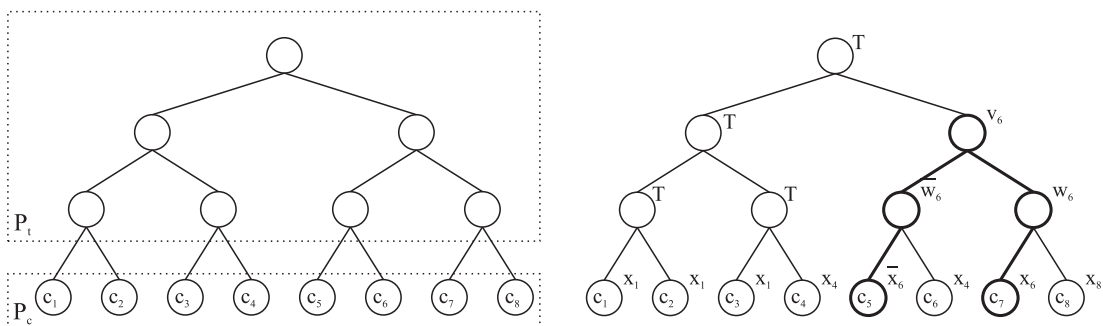

Figure 13: On the left: the dependency graph of the game $\mathcal{G}(\Phi_s)$. On the right: a coalition witnessing that $c_5$ and $c_7$ are playing in a conflicting way.

Finally, let us consider briefly the low-level implementation of the alternating Turing machine $M$. Existential states correspond to guesses, while universal states correspond to the recursive calls to $InCoalition_{\mathbf{x},JT}$, plus some further machinery for auxiliary computations. At each step, we have to encode on the worktape the two parameters $p$ and $st$ and the local variables, while $\mathbf{x}$, $JT$, the game $\mathcal{G}$ and the pre-computed constraint relations are on the input tape. Note that $p$ may be encoded by the logspace pointer to her position in the input tape, as well as any combined strategy $st_w$ may be encoded by a logspace pointer to its corresponding entry in the constraint relation associated with $w$. Similar considerations apply to the other local variables, e.g., the guessed combined strategy $st'$. Moreover, it is easy to check that all the computations performed by the function are feasible in logspace. Therefore $M$ is a logspace ATM, and the overall computation is in PTIME. (For a detailed description of such logspace ATM computations, we refer the interested reader to Gottlob et al., 2001; Gottlob, Leone, & Scarcello, 2002a). □

**Theorem 4.8** *Let $\mathcal{G}$ be an acyclic-hypergraph game that has small neighborhood or is in graphical normal form. Then, deciding whether $\mathcal{G}$ has strong Nash equilibria, i.e., $\mathcal{SNE}(\mathcal{G}) \neq \emptyset$ is NP-complete. Hardness holds even if $\mathcal{G}$ is in graphical normal form and has 3-bounded neighborhood.*

**Proof.** *Membership.* Given the game $\mathcal{G}$, we can guess a global strategy $\mathbf{x}$ and verify in polynomial time, by Lemma 4.7, that $\mathbf{x}$ is in fact a strong Nash equilibrium.

*Hardness.* The reduction is from SAT. Consider a Boolean formula in conjunctive normal form $\Phi = c_1 \wedge \ldots \wedge c_m$ over variables $X_1, \ldots, X_n$ and assume, w.l.o.g., that $m = 2^\ell$, for some $\ell > 0$. As a running example, consider the formula $\Phi_s = (X_1 \vee X_2) \wedge (X_1 \vee X_3) \wedge (X_1 \vee \neg X_4 \vee X_8) \wedge (X_4) \wedge (\neg X_5 \vee \neg X_6) \wedge (X_1 \vee X_4 \vee X_6) \wedge (X_6 \vee X_7) \wedge (X_8)$.

From $\Phi$, we build the following GNF game $\mathcal{G}(\Phi)$. The players are partitioned in two sets $P_c$ and $P_t$. The set $P_c$ contains exactly one player for each clause of $\Phi$, while players in $P_t$ are such that $G(\mathcal{G})$ is a complete binary tree, whose leaves are the players in $P_c$, as shown in Figure 13 for $\Phi_s$.

Players in $\mathcal{G}(\Phi)$ play actions corresponding to variables in $\Phi$ plus some further special actions. Intuitively, each player $c \in P_c$ may play a literal occurring in the clause she represents, while players in $P_t$ have to check that no pair of players $c_i, c_j \in P_c$ plays in a conflicting way, that is, plays complementary literals. To this end, the game rules are designed in such a way that players in $P_t$ may improve their payoffs if they are able to form a coalition proving that some pair of players are playing complementary literals. It is worthwhile noting that, in general, this situation cannot be detected by any single player, since the conflicting clauses may be very far from each other. For instance, in Figure 13, $c_5$ and $c_7$ play $\bar{x}_6$ and $x_6$, respectively, which is detected by a coalition involving their lowest common ancestor, say $p$, and the players of $P_t$ occurring in the two paths from $c_5$ and $c_7$ to $p$. We show that a global strategy is a strong Nash equilibrium of this game if and only if there is no such a disproving coalition. Indeed, in this case, there are no conflicting clauses and thus the formula $\Phi$ is satisfiable by setting to true all literals played by the clause players.

Formally, each player $c \in P_c$ may play either a special action $B$ (read: bad) or an action $x_i$ (resp. $\bar{x}_i$) called literal action, provided that $X_i$ is a variable occurring positively (resp. negatively) in the corresponding clause of $\Phi$. Each player $t \in P_t$ may play an action in $\{v_i, w_i, \bar{w}_i \mid X_i \text{ is a variable in } \Phi\} \cup \{T\}$, where $T$ can be read as "okay with me!"

We next describe the utility functions, given any global strategy $\mathbf{x}$.

A player $c \in P_c$ gets payoff 1 if she plays a literal action and her unique neighbor (i.e., her parent) plays $T$, or if she plays $B$ and her neighbor does not play $T$ **(C-i)**; otherwise, she gets payoff 0 **(C-ii)**.

For a player $t \in P_t$, the utility function $u_t$ is such that:

**(T-i)** $u_t(\mathbf{x}) = 2$ if $t$ plays $w_i$, her parent (if any) plays $w_i$ or $v_i$, none of her children plays $B$, and one of her children plays either $w_i$ or $x_i$ (depending on whether she is a leaf or not);

**(T-ii)** $u_t(\mathbf{x}) = 2$ if $t$ plays $\bar{w}_i$, her parent (if any) plays $\bar{w}_i$ or $v_i$, none of her children plays $B$, and one of her children plays either $\bar{w}_i$ or $\bar{x}_i$;

**(T-iii)** $u_t(\mathbf{x}) = 2$ if $t$ plays $v_i$, her parent (if any) plays $T$, and her children play either $x_i$ and $\bar{x}_i$ or $w_i$ and $\bar{w}_i$;

**(T-iv)** $u_t(\mathbf{x}) = 1$ if $t$ plays $T$;

**(T-v)** $u_t(\mathbf{x}) = 0$ in all the other cases.

Then, $\mathcal{G}(\Phi)$ has the following properties.

$P_1$ : *Let $\mathbf{x}$ be a global strategy for $\mathcal{G}(\Phi)$. Then, $\mathbf{x}$ is a Nash equilibrium if and only if all players in $P_t$ play $T$ in $\mathbf{x}$ and there is no player in $P_c$ playing $B$.*

If players in $P_t$ play $T$ and players in $P_c$ do not play $B$, then they get payoff 1 due to rules (T-iv) and (C-i). In this case, no player has an incentive to deviate, since by changing strategy she would get payoff 0 due to rules (T-v) and (C-ii).

For the other direction of the proof, let $\mathbf{x}$ be a Nash equilibrium and assume, by contradiction, that there is a player $c \in P_c$ choosing $B$. From (C-i) and (C-ii), it follows that some neighbor of $c$, say $t$, does not play $T$. However, this is impossible, because $t$ would get payoff 0 (from T-v) and could improve her payoff by playing $T$, contradicting the fact that $\mathbf{x}$ is a Nash equilibrium. Next, assume there exists a player in $P_t$ that does not play $T$, and let $t \in P_t$ be a player at the lowest possible level of the tree satisfying this assumption. It follows that the children of $t$ are clause players, for otherwise, by the choice of $t$, both of them should play $T$, and thus $t$ would get payoff 0 (T-V) and could improve to 1 by playing $T$. Therefore, $Neigh(t) \cap P_c \neq \emptyset$. Then, the only way for $t$ to get payoff greater than 0 comes from rule (T-iii), which means that her clause children play in a conflicting way, say $x_i$ and $\bar{x}_i$. However, since $t$ does not play $T$, they both get payoff 0 and thus could deviate from $\mathbf{x}$ by playing $B$ and getting payoff 1. Contradiction.

$P_2:$ *Let $\mathbf{x}$ be a Nash equilibrium for $\mathcal{G}(\Phi)$. Then, a coalition of players getting an incentive to deviate from $\mathbf{x}$ exists if and only if there are two clauses playing in $\mathbf{x}$ in a conflicting way.*

(If part.) Since $\mathbf{x}$ is a Nash equilibrium, from Property $P_1$, all players in $P_t$ play $T$ and get payoff 1. If there are two clauses, say $c_1$ and $c_2$, playing $x_i$ and $\bar{x}_i$, respectively, we may identify an improving coalition as follows: let $t$ be first common ancestor of $c_1$ and $c_2$, and let $P_1$ and $P_2$ be the sets of vertices (players) occurring in the paths from $t$ to $c_1$ and $c_2$, respectively. Then, let $t$ change to $v_i$, and all the players in $P_1$ (resp. $P_2$) change to $w_i$ (resp. $\bar{w}_i$) in $\mathbf{x}$ – see Figure 13. Then, from the game rules above, all players in the coalition $K = P_1 \cup P_2 \cup \{t\}$ get payoff 2, improving the payoff 1 that they get in $\mathbf{x}$.

(Only-if part.) Let $K$ be a coalition of players improving the Nash equilibrium $\mathbf{x}$. From property $P_1$, all players in $P_t \cap K$ should get payoff 2, as they get 1 in $\mathbf{x}$. Let $t \in K$ by the player in $P_t$ at the highest (close to the root) level in the tree, i.e., such that $parent(t)$, if any, does not belong to $K$. From (T-iii), the children of $t$ must play either $x_i$ and $\bar{x}_i$ or $w_i$ and $\bar{w}_i$, depending on whether they are or are not leaves of the tree. In the former case, we have identified two conflicting players and thus the property is immediately proved. Hence, let us investigate the latter one. Let $t'$ and $t''$ be the children of $t$ playing $w_i$ and $\bar{w}_i$, respectively. Since they do not play $T$, both $t'$ and $t''$ belong to $K$ and have to improve their payoff to 2. Therefore, a child of $t'$ must play $w_i$ and a child of $t''$ must play $\bar{w}_i$, according to (T-i) and (T-ii). Therefore, these players belong to $K$, too, and the same happen for some of their children. Eventually, a leaf descendant of $t'$ plays $x_i$ and a leaf descendant of $t''$ plays $\bar{x}_i$, qed.

The NP hardness of deciding the existence of a SNE follows from the following claim: $\Phi$ is satisfiable $\Leftrightarrow$ $\mathcal{G}(\Phi)$ admits a strong Nash equilibrium.

($\Rightarrow$) Assume $\Phi$ is satisfiable and take a satisfying assignment $\sigma$. Let $\mathbf{x}^\sigma$ be a global strategy such that: each player $c \in P_c$ plays any literal occurring in the clause $c$ that is true with respect to $\sigma$; and each player $t \in P_t$ plays $T$. From $P_1$, $\mathbf{x}^\sigma$ is a Nash equilibrium for $\mathcal{G}(\Phi)$. Moreover, by construction no pair of players choose conflicting actions in $\mathbf{x}^\sigma$. Hence, due to $P_2$, there exists no coalition of players getting an incentive by deviating from $\mathbf{x}^\sigma$, and thus $\mathbf{x}^\sigma$ is strong. ($\Leftarrow$) Let $\mathbf{x}$ be a strong Nash equilibrium for $\mathcal{G}(\Phi)$. Then, there is no coalition of players getting an incentive to deviate from $\mathbf{x}$. Due to $P_1$, no player in $P_c$ plays $B$, and due to $P_2$ no pair of players play in a conflicting way. Hence, $\sigma^{\mathbf{x}}$ witnesses that $\Phi$ is satisfiable. More precisely, it encodes an implicant of $\Phi$, that can be extended to a satisfying assignment choosing any truth value for all Boolean variables occurring in $\Phi$ not chosen by any player in $\mathbf{x}$. $\qquad\square$

## 5. Further Structurally Tractable Classes of Games

For strategic games, both the acyclic graph and the acyclic hypergraph assumptions are very severe restrictions, which are rather unlikely to apply in practical contexts. In this section, we prove that even more general and structurally complicated classes of games can be dealt with in an efficient way. We consider the notions of *treewidth* (Robertson & Seymour, 1986) and *hypertree width* (Gottlob et al., 2002b), which are the broadest known generalizations of graph and hypergraph acyclicity, respectively (Gottlob et al., 2000). We show that tractability results for acyclic games hold for these generalizations, too, and study the relationship between the two notions.

### 5.1 Hypertree Decompositions of Games

Let $H = (V, E)$ be a hypergraph. Denote by $vert(H)$ and $edges(H)$ the sets $V$ and $E$, respectively. Moreover, for any set of edges $E' \subseteq edges(H)$, let $vert(E') = \bigcup_{h \in E'} h$.

A *hypertree for a hypergraph* H is a triple $\langle T, \chi, \lambda \rangle$, where $T = (N, E)$ is a rooted tree, and $\chi$ and $\lambda$ are labeling functions which associate with each vertex $p \in N$ two sets $\chi(p) \subseteq vert(H)$ and $\lambda(p) \subseteq edges(H)$. If $T' = (N', E')$ is a subtree of $T$, we define $\chi(T') = \bigcup_{v \in N'} \chi(v)$. We denote the root of $T$ by $root(T)$. Moreover, for any $p \in N$, $T_p$ denotes the subtree of $T$ rooted at $p$.

**Definition 5.1 (Gottlob et al., 2002b)** A *hypertree decomposition* of a hypergraph H is a hypertree $HD = \langle T, \chi, \lambda \rangle$ for H, where $T = (N, E)$, which satisfies all the following conditions:

1. for each edge $h \in edges(H)$, there exists $p \in N$ such that $vert(h) \subseteq \chi(p)$ (we say that $p$ *covers* $h$);

2. for each vertex $Y \in vert(H)$, the set $\{p \in N \mid Y \in \chi(p)\}$ induces a (connected) subtree of $T$;

3. for each $p \in N$, $\chi(p) \subseteq vert(\lambda(p))$;

4. for each $p \in N$, $vert(\lambda(p)) \cap \chi(T_p) \subseteq \chi(p)$.

An edge $h \in edges(H)$ is *strongly covered* in $HD$ if there exists $p \in N$ such that $vert(h) \subseteq \chi(p)$ and $h \in \lambda(p)$. In this case, we say that $p$ strongly covers $h$. A hypertree decomposition $HD$ of hypergraph H is a *complete decomposition* of H if every edge of H is strongly covered in $HD$. The *width* of a hypertree decomposition $\langle T, \chi, \lambda \rangle$ is $max_{p \in vertices(T)}|\lambda(p)|$. The *hypertree width* $hw(H)$ of H is the minimum width over all its hypertree decompositions.

Note that for any constant $k$ checking whether a hypergraph has hypertree-width at most $k$ is feasible in polynomial time (Gottlob et al., 2002b).

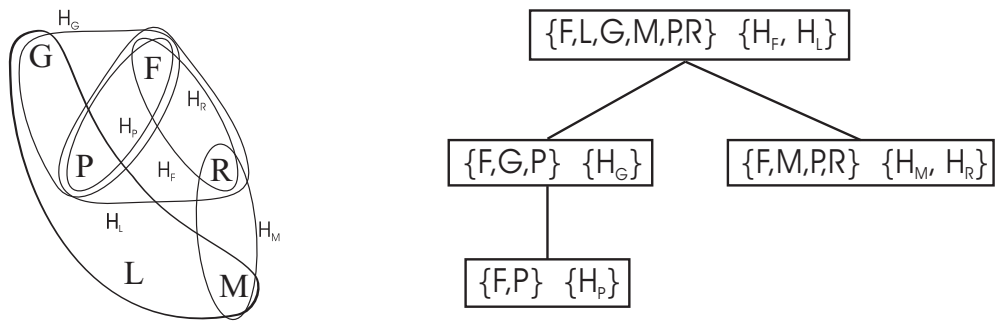

Figure 14: H(FRIENDS') and a hypertree decomposition for it.

Let $k > 0$ be a fixed constant. Then, we say that a game $\mathcal{G}$ has $k$-bounded hypertree width if the hypertree width of its associated hypergraph $H(\mathcal{G})$ is at most $k$. A hypertree decomposition of width at most $k$ (if any) can be computed in polynomial time.

Recall that the notion of bounded hypertree-width generalizes the notion of (hyper-graph) acyclicity. In particular, the class of acyclic-hypergraph games is precisely the class of games $\mathcal{G}$ whose hypergraph $H(\mathcal{G})$ has hypertree width 1.

**Example 5.2** Consider again the game FRIENDS in Example 2.1. Figure 5 shows on the left its associated hypergraph, and on the right a join tree for it. In fact, this join tree is a hypertree decomposition of width 1 for the hypergraph, where, for each vertex $p$, $\lambda(p)$ is the set of hyperedges reported in $p$ and $\chi(p)$ is the set of players occurring in these hyperedges.

For a more involved example, consider the extension FRIENDS' of FRIENDS where a new player *Laura* (short: $L$) joins the group. Laura would like to go with George to the cinema, and with Pauline and Mary to the opera. Figure 14 shows on the left the hyper-graph H(FRIENDS'). This hypergraph is not acyclic, but with a low degree of cyclicity. Indeed, its hypertree width is 2, as witnessed by the hypertree decomposition of width 2 shown on the right, in Figure 14. Here, for each vertex $p$ of the decomposition tree, the two sets denote its labels $\chi(p)$ and $\lambda(p)$, respectively. □

A class of games $C$ is said to have bounded hypertree-width if there is a finite $k$ such that, for each game $\mathcal{G} \in C$, $\mathcal{G}$ has $k$-bounded hypertree width. We next show that all the tractability results that hold for acyclic-hypergraph games holds for bounded hypertree-width games, as well.

**Theorem 5.3** *Deciding the existence of pure Nash equilibria, as well as computing a Nash equilibrium is feasible in polynomial time for all classes $\mathcal{C}$ of games having bounded hypertree-width and such that every game $\mathcal{G} \in \mathcal{C}$ has small neighborhood or is in graphical normal form.*

**Proof**. Let $\mathcal{C}$ be a class of games such that each game $\mathcal{G} \in \mathcal{C}$ has hypertree-width at most $k$, for some $k > 0$, and has small neighborhood or is in graphical normal form. Then, we can build the constraint satisfaction problem $CSP(\mathcal{G})$ in polynomial time, by Theorem 4.4.

Moreover, the hypertree width of $\mathcal{G}$ is at most $k$, and its hypergraph $H(\mathcal{G})$ is the same as the hypergraph H associated with $CSP(\mathcal{G})$. From results by Gottlob et al. (2001), it follows that $CSP(\mathcal{G})$ can be solved in polynomial time, which is equivalent to deciding the existence of Nash equilibria in polynomial time, by Theorem 4.3.

Constructively, we can compute in polynomial time a hypertree decomposition of H having width at most $k$, exploit this decomposition for building an equivalent acyclic problem (by putting together the constraints of players occurring in the same vertex of the decomposition tree), and finally solve this problem by using the algorithm shown in Figure 10. $\square$

As for acyclic-hypergraph games, this result can be immediately extended to the problem of computing a Pareto Nash equilibrium.

**Corollary 5.4** *Deciding the existence of Pareto Nash equilibria, as well as computing a Pareto Nash equilibrium for pure strategies is feasible in polynomial time for all classes $\mathcal{C}$ of games having bounded hypertree-width and such that every game $\mathcal{G} \in \mathcal{C}$ has small neighborhood or is in graphical normal form.*

## 5.2 Treewidth and Hypertree Width of Games

We next consider the treewidth of game structures. Recall that any game may be represented either by the primal graph or by the dual graph, as shown in Figure 5 for the game FRIENDS. Therefore, a first question is which graph is better as far as the identification of tractable classes of games is concerned. From the results by Gottlob et al. (2000), we know that the notion of bounded treewidth for the primal graph is generalized by the notion of bounded hypertree width, that is, looking at the hypertree width of the game hypergraph we may identify wider classes of tractable games. Moreover, from the results by Greco and Scarcello (2003), it follows that looking at the treewidth of the dependency graph is better than looking at the treewidth of the primal graph.[4]

We thus know that bounded treewidth for the primal graph is sufficient for ensuring game tractability. However, two questions are still to be answered, and will be the subject of this section:

1. Do tractability results for bounded treewidth for the primal graph extend to the wider class of games having bounded treewidth for the dependency graph?

2. What is the relationship between bounded treewidth for the dependency graph and bounded hypertree width of the game hypergraph?

**Definition 5.5 (Robertson & Seymour, 1986)** A *tree decomposition* of a graph $G = (V, E)$ is a pair $\langle T, \chi \rangle$, where $T = (N, F)$ is a tree, and $\chi$ is a labeling function assigning to each vertex $p \in N$ a set of vertices $\chi(p) \subseteq V$, such that the following conditions are satisfied:

(1) for each vertex $b$ of $G$, there exists $p \in N$ such that $b \in \chi(p)$;

---

4. In fact, this result is on relationship between primal graph and incidence graph. However, it is easy to see that, for games, the treewidth of the incidence graph is the same as the treewidth of the dependency graph.

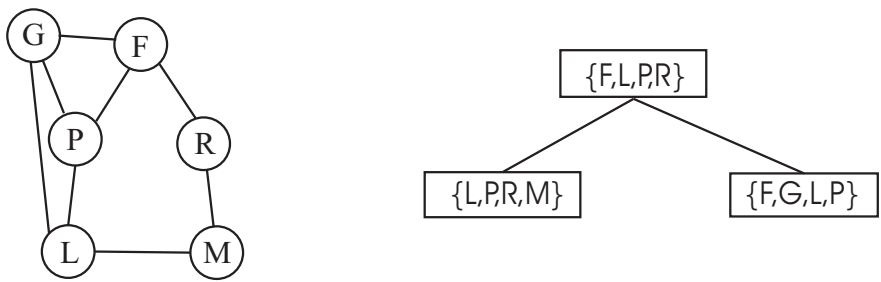

Figure 15: G(FRIENDS') and a tree decomposition for it.

(2) for each edge $\{b, d\} \in E$, there exists $p \in N$ such that $\{b, d\} \subseteq \chi(p)$;

(3) for each vertex $b$ of $G$, the set $\{p \in N \mid b \in \chi(p)\}$ induces a connected subtree of $T$.

Note that Condition 1 is subsumed by Condition 2 for graphs without isolated vertices. The *width* of the tree decomposition $\langle T, \chi \rangle$ is $\max_{p \in N} |\chi(p) - 1|$. The *treewidth* of $G$ is the minimum width over all its tree decompositions. This notion generalizes graph acyclicity, as the acyclic graphs are precisely those graphs having treewidth 1. [5]

**Example 5.6** Consider again the game FRIENDS' introduced in the Example 5.2. Figure 15 shows on the left the cyclic dependency graph G(FRIENDS'), and on the right a tree decomposition of width 3 of this graph. □

Let $k > 0$ be a fixed integer, and let $\mathcal{G}$ be a game. We say that a game $\mathcal{G}$ has $k$-bounded treewidth if the treewidth of its dependency graph G($\mathcal{G}$) is at most $k$. Recall that, given a graph G, computing a tree-decomposition of width at most $k$ of G (if any) is feasible in polynomial (actually, linear) time (Bodlaender, 1997).

We next prove an interesting graph-theoretic result to shed some light on the different possible representations of game structures: every class of games having bounded treewidth has bounded hypertree width, too. We remark that the previous results on the relationship between treewidth and hypertree width described in the literature (e.g. Gottlob et al., 2000) cannot be used here. Indeed, they deal with the primal graph and the dual graph representations, while we are now interested in the dependency graph, which is more effective than the primal graph, and somehow incomparable with the (optimal) dual graph. A detailed comparison of these two latter notions is reported by Greco and Scarcello (2003).

**Theorem 5.7** For each game $\mathcal{G}$, hypertreewidth(H($\mathcal{G}$)) $\leq$ treewidth(G($\mathcal{G}$)) + 1.

**Proof**. Let $TD$ be a tree decomposition of G($\mathcal{G}$) and let $k - 1$ be its width, that is, the largest label of the vertices of $TD$ contains $k$ players. Then, we show that there is a hypertree decomposition of H($\mathcal{G}$) having width $k$. Recall that H($\mathcal{G}$) contains, for each player $p$, the characteristic edge $\mathrm{H}(p) = \{p\} \cup Neigh(p)$. Let $HD = \langle T, \chi, \lambda \rangle$ be a hypertree such that:

- the tree $T$ has the same form as the decomposition tree $TD$, i.e., there is a tree isomorphism $\delta : vert(T) \longrightarrow vert(TD)$ between $T$ and $TD$;

- for each vertex $v \in T$, $\lambda(v) = \{\mathrm{H}(p) \mid p \in \delta(v)\}$, i.e., $\lambda(v)$ contains the characteristic edge of each player occurring in the vertex of the tree decomposition corresponding to $v$;

- $\chi(v)$ is the set of all vertices occurring in the edges in $\lambda(v)$, i.e., contains all players in $\delta(v)$ and their neighbors.

---

5. Observe that the "$-1$" in the definition of treewidth has been introduced in order to get this correspondence with acyclic graphs, as 2 is the minimum cardinality for the largest label in any tree decomposition.

Note that the width of $HD$ is $k$, as it is determined by the largest $\lambda$ label, which contains the same number of elements as the largest label in $TD$.

We claim that $HD$ is a hypertree decomposition of $\mathrm{H}(\mathcal{G})$. Consider the four conditions in Definition 5.1: Conditions 3 and 4 are trivially satisfied because, for each vertex $v$, $\chi(v) = vert(\lambda(v))$, by construction. Condition 1 is guaranteed by the fact that $TD$ satisfies its corresponding Conditions 1 and 2. We next show that Condition 2, i.e., the connectedness condition, holds, too.

Let $v_1$ and $v_2$ be two vertices of $T$ such that there exists $p \in \chi(v_1) \cap \chi(v_2)$. Let $v_1' = \delta(v_1)$ and $v_2' = \delta(v_2)$ be the sets of vertices in the tree decomposition $TD$ corresponding to $v_1$ and $v_2$, respectively. Since $p \in \chi(v_1)$ and $p \in \chi(v_2)$, there are two players $p_1$ and $p_2$ such that (i) $\mathrm{H}(p_1) \in \lambda(v_1)$ and $p \in \mathrm{H}(p_1)$, and (ii) $\mathrm{H}(p_2) \in \lambda(v_2)$ and $p \in \mathrm{H}(p_2)$. Then, by construction, $p_1 \in v_1'$ and $p_2 \in v_2'$ (see Figure 16).

We claim that, for each vertex $v$ in the unique path connecting $v_1$ and $v_2$ in $T$ (denoted by $v_1 \rightsquigarrow v_2$), $\lambda(v)$ contains a player from the set $\{p_1, p_2, p\}$, which entails that $p \in \chi(v)$ and hence that Condition 2 is satisfied by $HD$. This is equivalent to claim, on the tree decomposition $TD$, that each vertex $v'$ in the path $v_1' \rightsquigarrow v_2'$ contains a player in $\{p_1, p_2, p\}$.

If both $v_1'$ and $v_2'$ contain $p$, then the claim trivially holds because all the vertices in the path $v_1' \rightsquigarrow v_2'$ must contain $p$, from Condition 3 of tree decompositions (the connectedness condition).

Hence, let us assume that $v_1'$ does not contain $p$. Since $p \in H(p_1)$, this means that $p$ is a neighbor of $p_1$ and thus there exists a vertex of $TD$, say $v_3' \neq v_1'$, whose labeling contains both $p$ and $p_1$. Assume now that $v_2'$ contains $p$. Figure 16.1 shows the path comprising vertices $v_1'$, $v_3'$, and $v_2'$ — notice that $v_1'$ cannot be in the path $v_3' \rightsquigarrow v_2'$, otherwise it should contain $p$ as well. The result follows by observing that, again from Condition 3 of tree decompositions, all vertices in the path $v_1' \rightsquigarrow v_3'$ must contain $p_1$, and all vertices in the path $v_3' \rightsquigarrow v_2'$ must contain $p$. Similarly, assume that $v_2'$ does not contain $p$. In this case, since $p$ is a neighbor of $p_2$ (recall the above discussion for $p_1$), there is a vertex $v_4'$ in $TD$ whose labeling contains both $p_2$ and $p$. Figure 16.2 shows how these vertices should look like in the tree decomposition $TD$. Then, the result follows by observing that all vertices in the path $v_1' \rightsquigarrow v_3'$ contains $p_1$, all vertices in the path $v_3' \rightsquigarrow v_4'$ contains $p$, and all vertices in the path $v_4' \rightsquigarrow v_2'$ contains $p_2$. $\qquad \square$

We next show that the converse does not hold, that is, there are classes of games having bounded hypertree width, but unbounded treewidth. That is, the technique based on the hypertree width of the game hypergraph is more effective than the corresponding technique based on the treewidth of the dependency graph, because it allows us to identify strictly broader classes of tractable games.

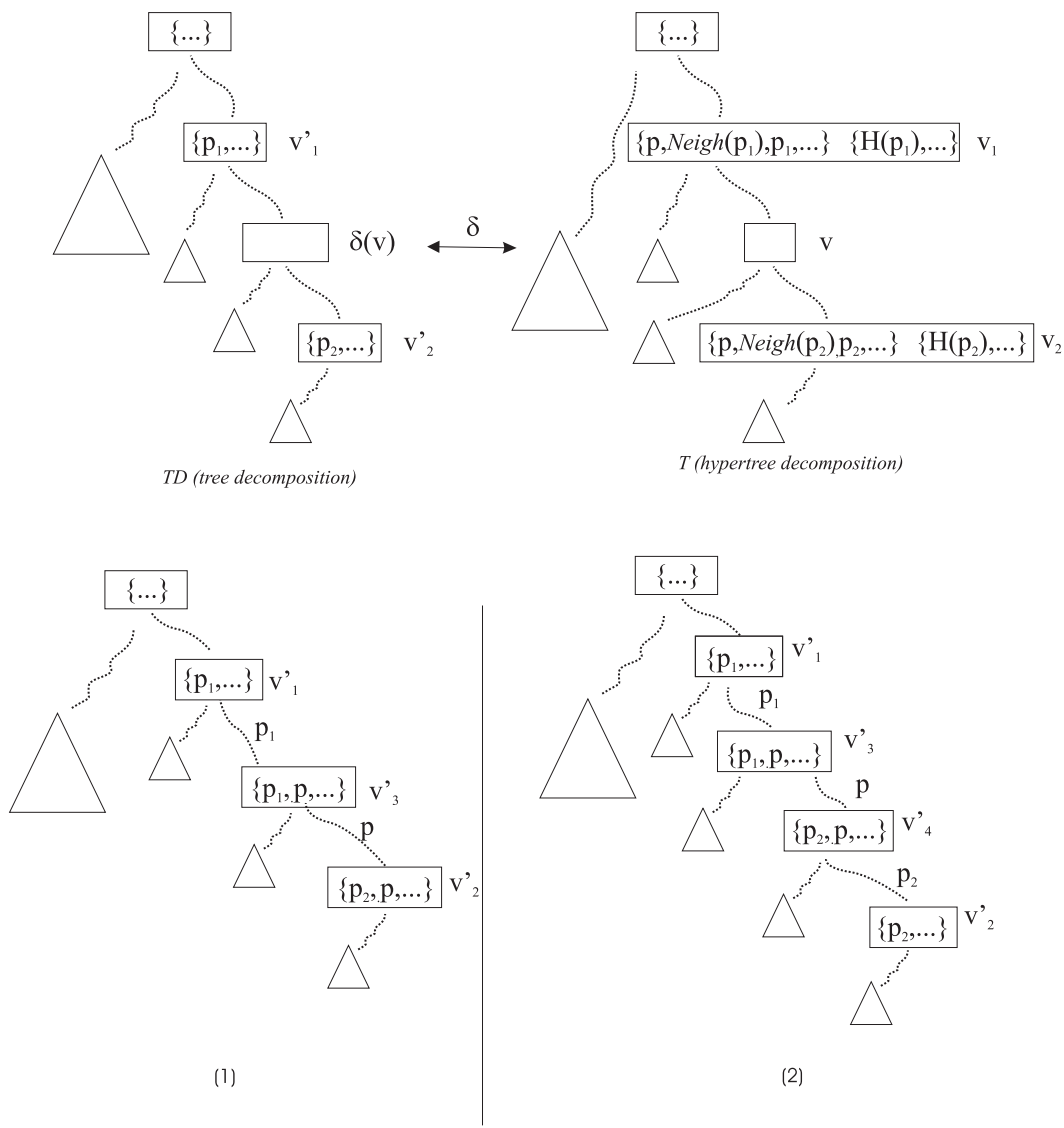

Figure 16: Schema of the reduction in the proof of Theorem 5.7.

**Theorem 5.8** *There are classes C of games having hypertree width 1 but unbounded treewidth, i.e., such that, for any finite $k > 0$, there is a game $\mathcal{G} \in C$ such that the treewidth of $G$ is not bounded by $k$.*

**Proof**. Take the class of all games where every player depends on all other players. For every such game $\mathcal{G}$, $H(\mathcal{G})$ is acyclic and thus its hypertree width is 1, while $G(\mathcal{G})$ is a clique containing all players and its treewidth is the number of players minus 1. $\square$

From Theorem 5.3, Corollary 5.4, and Theorem 5.7, we immediately get the following tractability results for bounded treewidth games.

**Corollary 5.9** *Deciding the existence of pure (Pareto) Nash equilibria, as well as computing a pure (Pareto) Nash equilibrium is feasible in polynomial time for all classes $\mathcal{C}$ of games having bounded treewidth and such that every game $\mathcal{G} \in \mathcal{C}$ has small neighborhood or is in graphical normal form. Moreover, all Nash equilibria of such games can be computed in time polynomial in the combined size of input and output.*

## 6. Parallel Complexity of Easy Games

In this section, we show that dealing with Nash equilibria for games with good structural properties is not only tractable but also parallelizable. More precisely, we show that deciding the existence of Nash equilibria for graphical games where the player interactions has a low degree of cyclicity is complete for the class LOGCFL. Also, we show that computing such an equilibrium belongs to the functional version of LOGCFL.

The complexity class LOGCFL consists of all decision problems that are logspace reducible to a context-free language. In order to prove the following theorem, we exploit a characterization of LOGCFL in terms of circuits.

We recall that a *Boolean circuit* $G_n$ with $n$ inputs is a finite directed acyclic graph whose nodes are called *gates* and are labeled as follows. Gates of fan-in (indegree) zero are called *circuit input gates* and are labeled from the set $\{false, true, z_1, z_2, \ldots, z_n, \neg z_1, \neg z_2, \ldots, \neg z_n\}$. All other gates are labeled either AND, OR, or NOT. The fan-in of gates labeled NOT must be one. The unique node with fan-out (outdegree) zero is called *output gate*. The evaluation of $G_n$ on input string $w$ of length $n$ is defined in the standard way. In particular, any input gate $g$ labeled by $z_i$ (resp. $\neg z_i$) gets value *true* (resp. *false*) if the $i$th bit of $w$ is 1 (resp. 0); otherwise, $g$ gets value *false* (resp. *true*).

A Boolean circuit is thus given as a triple $(N, A, label)$, where $N$ is the set of nodes (gates), $A$ is the set of arcs, and *label* is the labeling of the nodes as described.

The *depth* of a Boolean circuit $G$ is the length of a longest path in $G$ from a circuit input gate to the output gate of $G$. The size $S(G)$ of $G$ is the number of gates (including input-gates) in $G$.

A family $\mathcal{G}$ of Boolean circuits is a sequence $(G_0, G_1, G_2, \ldots)$, where the $n$th circuit $G_n$ has $n$ inputs. Such a family is *logspace-uniform* if there exists a logspace Turing machine which, on the input string containing $n$ bits 1, outputs the circuit $G_n$. Note that the size of the $n$th circuit $G_n$ of a logspace-uniform family $\mathcal{G}$ is polynomial in $n$. Intuitively, this uniformity condition is crucial in characterizations of low parallel complexity classes in terms of circuits, because hidden inherent sequentialities in the circuit construction process must be avoided. In fact, the cicuits which serve as parallel devices for evaluating input strings of length $n$, and which must be constructed for each $n$ separately, should be constructible in parallel themselves. This is assured by requiring logspace uniformity, because LOGSPACE is a highly parallelizable complexity class contained in LOGCFL.

The language $L$ accepted by a family $\mathcal{G}$ of circuits is defined as follows: $L = \bigcup_{n \geq 0} L_n$, where $L_n$ is the set of input strings accepted by the $n$th member $G_n$ of the family. An input string $w$ of length $n$ is accepted by the circuit $G_n$ if $G_n$ evaluates to *true* on input $w$.

A family $\mathcal{G}$ of Boolean circuits has *bounded fan-in* if there exists a constant $c$ such that each gate of each member $G_n$ of $\mathcal{G}$ has its fan-in bounded by $c$.

A family $\mathcal{G}$ of Boolean circuits is *semi-unbounded* if the following two conditions are met:

- All circuits of $\mathcal{G}$ involve as non-leaves only AND and OR gates, but no NOT gates (negation may thus only occur at the circuit input gates); and

- there is a constant $c$ such that each AND gate of any member $G_n$ of $\mathcal{G}$ has its fan-in bounded by $c$ (the OR gates may have unbounded fan-in).

For $i \geq 1$, $AC_i$ denotes the class of all languages recognized by logspace-uniform families of Boolean circuits of depth $O(\log^i n)$.

For $i \geq 1$, $NC_i$ denotes the class of all languages recognized by logspace-uniform families of Boolean circuits of depth $O(\log^i n)$ having bounded fan-in.

For $i \geq 1$, $SAC^i$ denotes the class of all languages recognized by semi-unbounded logspace-uniform families of Boolean circuits of depth $O(\log^i n)$.

Venkateswaran (1991) proved the following important relationship between LOGCFL and the semi-unbounded circuits:

$$LOGCFL = SAC^1.$$

Since $LOGCFL = SAC^1 \subseteq AC_1 \subseteq NC_2$, the problems in LOGCFL are all highly parallelizable. In fact, each problem in LOGCFL is solvable in logarithmic time by a concurrent-read concurrent-write parallel random access machine (CRCW PRAM) with a polynomial number of processors, or in $\log^2$-time by an exclusive-read exclusive-write PRAM (EREW PRAM) with a polynomial number of processors (Johnson, 1990).

We next show that the evaluation problem of $SAC^1$ circuits can be transformed in logspace into the considered Nash equilibrium existence problems.

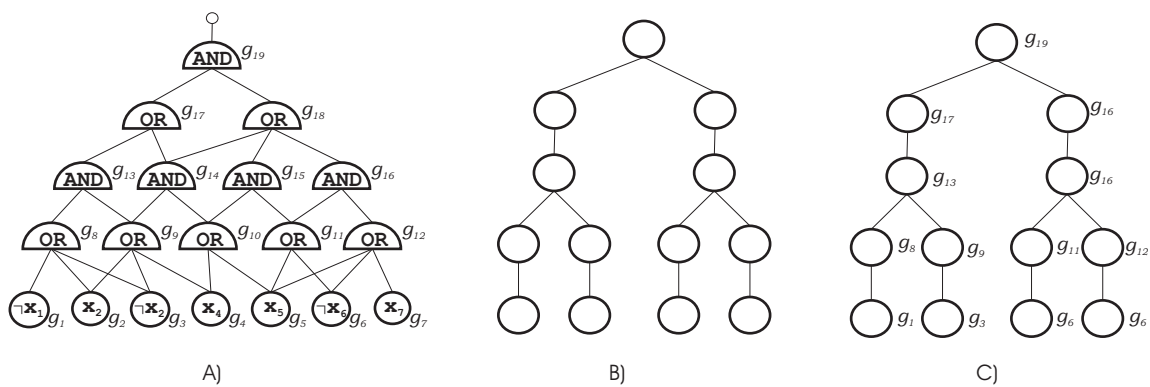

Figure 17: (A) A normalized circuit, (B)its skeleton tree, (C) a labeling corresponding to a proof tree.

**Theorem 6.1** *The existence problem for pure Nash equilibria is* LOGCFL-*complete for the following classes of strategic games in graphical normal form: acyclic-graph games, acyclic-hypergraph games, games of bounded treewidth, and games of bounded hypertree-width.*

**Proof.** It is sufficient to show membership for bounded hypertree width (the largest of the 4 classes) and hardness for acyclic-graph games (the smallest one).

*Membership.* The Nash equilibrium existence problem for NF games of bounded hypertree width is in LOGCFL because, as shown in Section 4, this problem can be transformed in logspace into a CSP of bounded hypertree width, and, as shown by Gottlob et al. (2001), checking satisfiability of the latter is in LOGCFL. (Recall that LOGCFL is closed under logspace reductions.)

*Hardness.* We assume that a logspace-uniform family $\mathcal{C} = \{G_1, G_2, \ldots\}$ of SAC[1] circuits is given, and we prove that the problem of checking whether a binary string $w$ is accepted by $\mathcal{C}$ can be translated in logspace into an acyclic Nash equilibrium problem in NF.

On input $w$, compute in logspace the appropriate circuit $C = G_{|w|}$. As shown by Gottlob et al. (2001), this circuit can be transformed in logspace into an equivalent normalized circuit $C'$ which is stratified and strictly alternating (see Figure 17 (A)), and a tree-shaped proof skeleton $SKEL$ (see Figure 17 (B)) which encompasses the common structure of all possible *proof trees* for $(C', w)$. A proof tree is a subtree of $C'$ of gates having value 1 witnessing that $C'$ accepts $w$. Each proof tree corresponds to an appropriate labeling of SKEL with gates from $C'$ (for example, the labeling shown in Figure 17 (C)). A labeling is correct if the root of SKEL is labeled with the output gate of $C'$, each AND gate labeled $g$ has two children that are labeled with the input gates to $g$, each OR node of SKEL labeled $g$ has one child labeled with some input gate to $g$, and each leaf of $T$ is labeled with an input gate to $C'$ whose output to the next higher level is 1. $C'$ accepts $w$ if and only if there exists a proof tree for $(C', w)$, and thus if there exists a correct labeling of $SKEL$.

Build a strategic game $\mathcal{G}$ from $(C', w)$ and SKEL as follows. The set of players consists of all vertices $V$ of $SKEL$ plus two special players $\alpha$ and $\beta$. The possible actions for the players in $V$ are pairs $(g, t)$, where $g$ is a gate and $t$ is a truth value in $\{true, false\}$. The utilities for players in $V$ are given as follows.

1. The utility of each leaf $u$ of $SKEL$ only depends on its action and is 1 if it plays an input gate $g$ of $C'$ and if $g$ is associated with the constant *true*, or $g$ is a $\neg$ gate and corresponds to an input bit 0 of the string $w$, or $g$ is not a $\neg$ gate and corresponds to an input bit 1 of $w$. Otherwise, the utility of $u$ is 0.

2. Each non leaf vertex $p \in V$ gets payoff 1, if it plays an action $(g, true)$, and if either $p$ is an OR vertex and the unique child of $p$ in $SKEL$ takes action $(g', true)$, and $g'$ is a child of the gate $g$ in $C'$, or $p$ is an AND vertex and the unique children of $p$ in $SKEL$ take actions $(g', true)$, and $(g'', true)$, respectively, where $g'$ and $g''$ are the children of the gate $g$ in $C'$.

3. Each non leaf vertex $p \in V$ gets payoff 1, if it plays an action an action $(g, false)$, and if either $p$ is an OR vertex and the unique child of $p$ in $SKEL$ takes action $(g', false)$, and $g'$ is a child of $g$ in $C'$, or $p$ is an AND vertex and the unique children of $p$ in $SKEL$ take actions $(g', t')$, and $(g'', t'')$, respectively, where $g'$ and $g''$ are the children of $g$ in $C'$ and $t' \wedge t'' = false$.

4. In all other cases, all actions of a non leaf vertex $p \in V$ have utility $-1$.

According to what we have defined so far, it is easy to see that every Nash equilibrium of the game corresponds to a labeling of $SKEL$ by assigning each player of $V$ the gate $g$ of its respective action $(g, t)$. In particular, the root node $r$ is forced to be labeled with the output gate $g^*$ of $C'$, and the action played by $r$ is $(g^*, true)$ if and only if the particular labeling is a proof tree and $(g^*, false)$ otherwise.

It remains to define the actions and utilities for the special players $\alpha$ and $\beta$. Their intuitive role is to "kill" all those equilibria which do not correspond to a proof three i.e., those where the root vertex plays $(g^*, false)$. The possible actions are $\{ok, head, tail\}$ for $\alpha$, and $\{head, tail\}$ for $\beta$. The strategies where $\alpha$ plays $ok$ have utility 1 for $\alpha$ if the root vertex $r$ of $SKEL$ plays $(g^*, true)$) and utility 0 otherwise. Strategies where $\alpha$ plays $head$ (resp., $tail$) have utility 1 for $\alpha$ if $r$ plays $(g^*, false)$ and if $\beta$ plays $tail$ (resp., $head$), and 0 otherwise. Thus, in case $r$ plays $(g^*, false)$, player $\alpha$ tries to play the opposite of player $\beta$. The strategies where $\beta$ plays $head$ (resp., $tail$) have utility 1 for player $\beta$ if $\alpha$ plays the same action, and 0 otherwise.

Therefore, in case $C'$ outputs 0 on input $w$, $r$ plays $(g^*, false)$, and thus $\alpha$ tries to play opposite to $\beta$ while $\beta$ tries to mimic $\alpha$. This is a classical non-equilibrium situation. In summary, each Nash equilibrium of $\mathcal{G}$ corresponds to a proof tree for $(C', w)$. Note also that $G(\mathcal{G})$ is acyclic, and that the construction of $\mathcal{G}$ from $(C', w)$ can be done in logspace. □

Note that, by Definition 2.2, a Pareto Nash equilibrium exists if and only if a Nash equilibrium exists.

**Corollary 6.2** *The existence problem for pure Pareto Nash equilibria is* LOGCFL-*complete for the following classes of strategic games in graphical normal form: acyclic-graph games, acyclic-hypergraph games, games of bounded treewidth, and games of bounded hypertree-width.*

Finally, as far as computation of Nash equilibria is concerned, the following corollary follows from the above result and from a result by Gottlob at al. (2002a), stating that witnesses (i.e., proof trees) of LOGCFL decision problems can be computed in functional LOGCFL (i.e., in logspace with an oracle in LOGCFL, or, equivalently, by using SAC[1] circuits.

**Corollary 6.3** *For the classes of games mentioned in Theorem 6.1, the computation of a single pure Nash equilibria can be done in functional LOGCFL, and is therefore in the parallel complexity class* $NC_2$.

## 7. Conclusion

In this paper we have determined the precise complexity of pure Nash equilibria in strategic games. As depicted in Figure 2, our study proceeded along three directions: representation issues, structural properties of player interactions, and different notions of equilibria. Indeed, besides "plain" Nash equilibria, we considered Pareto and Strong Nash equilibria, where we look for Nash equilibria that are not dominated by any other Nash equilibrium, or for profiles where no possible coalition of players may improve the payoffs of all its members, respectively.

It turns out that, apart from the simple case of standard normal form, deciding the existence of Nash equilibria is an intractable problem (unless PTIME = NP), if there is no restriction on the relationships among players. Interestingly, for Strong Nash Equilibria, this problem is located at the second level of the polynomial hierarchy, and gives us a fresh game-theoretic view of the class $\Sigma_2^P$, as the class of problems whose positive instances are characterized by a coalition of players who cooperate to provide an equilibrium, and win against any other disjoint coalition, which fails in trying to improve the utility for all of its players.

However, this paper is not just a collection of bad news. Rather, a central goal was to single out large classes of strategic games where detecting Nash equilibria is a tractable problem. In particular, while early studies in game theory mainly focused on games with a small number of players (e.g., the traditional two-player framework), we are interested here in large population games, too. In such cases, adopting the standard normal form is clearly impractical as, for each player, one should specify her payoffs for any combination of choices of all players in the game. We thus considered a different representation for these games, known in literature as *graphical games* (Kearns et al., 2001b), where the payoffs of each player $p$ are functions of $p$'s neighbors only, that is, $p$'s utility function depends only on those players $p$ is directly interested in. These relationships among players may be represented as a graph or, more faithfully, as a hypergraph. We showed that, if utility functions are represented as tables (graphical normal form) and the game structure is acyclic or has a low degree of cyclicity (i.e., it has bounded hypertree width), then deciding the existence of a Nash equilibrium and possibly computing it is feasible in polynomial time. These results complement those obtained for graphical games in the mixed Nash equilibria framework (e.g. Kearns et al., 2001b; Kearns & Mansour, 2002). Moreover, in the case of quasi-acyclic structures, we were also able to extend tractability to classes of games where utility functions are given implicitly (as in the general form), provided that each player has a small number of neighbors with not too many available actions.

This paper sheds light on the sources of complexity of finding pure Nash equilibria in strategic games, and, in particular, on the roles played by game representations and game structures. It is worthwhile noting that these aspects of game theory have received a renewed deal of attention recently. For instance, see Papadimitriou (2004) for a recent work on the complexity of pure Nash equilibria in some particular classes of games, and the various contributions on different kinds of concise game representations (e.g. Koller & Milch, 2001; Vickrey, 2002; Kearns et al., 2001b; Leyton-Brown & Tennenholtz, 2003; Gal & Pfeffer, 2004; Kearns & Mansour, 2002).

We recall that a preliminary version of the present work has been presented at the 9th ACM Conference on Theoretical Aspects of Rationality and Knowledge (TARK'03). Since then, our results have been extended along different directions. In particular, Alvarez et al. (2005) considered a further version of general form games, called games in *implicit form*, where also payoff values are given in a succinct way. They showed that, for such games, the complexity of deciding the existence of pure Nash equilibria increases from the first level to the second level of the polynomial hierarchy. We point out that our general form is slightly different from the general form adopted in the above mentioned paper, and some confusion may arise by reading their citation of the results presented in our TARK'03 paper (whose full version is the present paper). In their terminology, our Turing-machine encoding of payoff functions in general form games should be classified as non-uniform, with a uniform time-bound. However, apart from such subtle technical issues, some of their results on general form games with non implicit actions are very similar to ours, but their contributions focus on games with a large number of actions, while our hardness results hold even for games with a fixed number of actions and payoff levels. Moreover, while we show that hardness holds even for acyclic games, they did not consider any restriction on player interactions. Observe that their results may be immediately strengthened, given that from our proofs about GNF games with arbitrary player interactions it follows that NP-hardness holds even for constant-time utility functions (as discussed in Remark 3.9).

Another line of research studies games where the computation of *any* Nash equilibrium is not satisfactory, and one is rather interested in equilibria that satisfy some additional requirements (e.g., the best social welfare). Greco and Scarcello (2004) proved that deciding the existence of such pure Nash equilibria, called *constrained Nash equilibria*, is intractable even for very simple requirements. However, they were also able to identify some restrictions (for player interactions and requirements) making both the existence and the computation problems easy. Recent contributions on this subject (on both pure and mixed Nash equilibria) have been done by Schoenebeck at al. (2005) and Greco and Scarcello (2005).

Finally, we observe that there is an interesting connection among strong Nash equilibria and some equilibria studied in cooperative/coalitional game theory (e.g. Mas-Colell, Whinston, & Green, 1995). In this framework, for each subset $K$ of the players, we are given the utility that players in $K$ may get, if they cooperate together. The *core* of a game is the set of profiles **x** such that there is no subset of players that may improve their utilities by forming their own coalition, deviating from **x** (Gillies, 1953). Recently, Conitzer and Sandholm (2003a) proposed a concise representation for coalition utilities, and showed that determining whether the core of such a game is nonempty is NP-hard. An interesting future work may concern a detailed study of the complexity of these coalitional games, possibly exploiting suitable notions of quasi-acyclic structures for identifying relevant tractable classes.

## Acknowledgments

Part of this work has been published in preliminary form in the Proceedings of the 9th ACM Conference on Theoretical Aspects of Rationality and Knowledge (TARK'03).

Georg Gottlob's work was supported by the Austrian Science Fund (FWF) under project *Nr. P17222-N04 Complementary Approaches to Constraint Satisfaction*, and by the *GAMES Network of Excellence of the EU*.

We thank the anonymous referees and Tuomas Sandholm for their very useful comments.

## References

Alvarez, C., Gabarro, J., & Serna, M. (2005). Pure Nash equilibria in games with a large number of actions. *Electronic Colloquium on Computational Complexity, Report TR05-031*.

Aumann, R. (1959). Accetable points in general cooperative n-person games. *Contribution to the Theory of Games, IV*.

Aumann, R. (1985). What is game theory trying to accomplish?. *Frontiers of Economics*, 28–76.

Beeri, C., Fagin, R., Maier, D., & Yannakakis, M. (1983). On the desirability of acyclic database schemes. *Journal of the ACM, 30(3)*, 479–513.

Bodlaender, H. (1997). Treewidth: Algorithmic techniques and results. In *Proc. of the 22nd International Symposium on Mathematical Foundations of Computer Science (MFCS'97)*, pp. 19–36, Bratislava, Slovakia.

Chandra, A., Kozen, D., & Stockmeyer, L. (1981). Alternation. *Journal of the ACM, 28(1)*, 114–133.

Conitzer, V., & Sandholm, T. (2003a). Complexity of determining nonemptiness of the core. In *Proc. of the 18th International Joint Conference on Artificial Intelligence (IJCAI'03)*, pp. 613–618, Acapulco, Mexico.

Conitzer, V., & Sandholm, T. (2003b). Complexity results about nash equilibria. In *Proc. of the 18th International Joint Conference on Artificial Intelligence (IJCAI'03)*, pp. 765–771, Acapulco, Mexico.

Deng, X., Papadimitriou, C., & Safra, S. (2002). On the complexity of equilibria. In *Proc. of the 34th Annual ACM Symposium on Theory of Computing (STOC'02)*, pp. 67–71, Montreal, Canada.

Downey, R., & Fellows, M. (1995). Fixed-parameter tractability and completeness i: Basic results. *SIAM Journal on Computing, 24(4)*, 873–921.

Fabrikant, A., Papadimitriou, C., & Talwar, K. (2004). The complexity of pure nash equilibria. In *Proc. of the 36th Annual ACM Symposium on Theory of Computing (STOC'04)*, pp. 604–612, Chicago, IL, USA.

Fagin, R. (1983). Degrees of acyclicity for hypergraphs and relational database schemes. *Journal of the ACM, 30(3)*, 514–550.

Fotakis, D., Kontogiannis, S., Koutsoupias, E., Mavronicolas, M., & Spirakis, P. (2002). The structure and complexity of nash equilibria for a selfish routing game. In *Proc. of the 29th International Colloquium on Automata, Languages and Programming (ICALP'02)*, pp. 123–134, Malaga, Spain.

Gal, Y., & Pfeffer, A. (2004). Reasoning about rationality and beliefs. In *Proc. of the 3rd International Joint Conference on Autonomous Agents and Multiagent Systems (AAMAS'04)*, pp. 774–781, New York, NY, USA.

Garey, M., & Johnson, D. (1979). *Computers and Intractability. A Guide to the Theory of NP-completeness*. Freeman and Comp., NY, USA.

Gilboa, I., & Zemel, E. (1989). Nash and correlated equilibria: Some complexity considerations. *Games and Economic Behaviour, 1*, 80–93.

Gillies, D. (1953). Some theorems on n-person games. *PhD thesis, Princeton, Dept. of Mathematics.*

Gottlob, G., Leone, N., & Scarcello, S. (2000). A comparison of structural csp decomposition methods. *Artificial Intelligence, 124(2)*, 243–282.

Gottlob, G., Leone, N., & Scarcello, S. (2001). The complexity of acyclic conjunctive queries. *Journal of the ACM, 48(3)*, 431–498.

Gottlob, G., Leone, N., & Scarcello, S. (2002a). Computing logcfl certificates. *Theoretical Computer Science, 270(1-2)*, 761–777.

Gottlob, G., Leone, N., & Scarcello, S. (2002b). Hypertree decompositions and tractable queries. *Journal of Computer and System Sciences, 63(3)*, 579–627.

Greco, G., & Scarcello, S. (2003). Non-binary constraints and optimal dual-graph representations. In *Proc. of the 18th International Joint Conference on Artificial Intelligence (IJCAI'03)*, pp. 227–232, Acapulco, Mexico.

Greco, G., & Scarcello, S. (2004). Constrained Pure Nash Equilibria in Graphical Games. In *Proc. of the 16th Eureopean Conference on Artificial Intelligence (ECAI'04)*, pp. 181–185, Valencia, Spain.

Greco, G., & Scarcello, S. (2005). Bounding the Uncertainty of Graphical Games: The Complexity of Simple Requirements, Pareto and Strong Nash Equilibria. to appear In *Proc. of the 21st Conference in Uncertainty in Artificial Intelligence (UAI'05)*, Edinburgh, Scotland.

Johnson, D. (1990). A catalog of complexity classes. *Handbook of Theoretical Computer Science, Volume A: Algorithms and Complexity*, 67–161.

Johnson, D., Papadimitriou, C., & Yannakakis, M. (1998). How easy is local search?. *Journal of Computer and System Sciences, 37*, 79–100.

Kearns, M., Littman, M., & Singh, S. (2001a). An efficient exact algorithm for singly connected graphical games. In *Proc. of the 14th International Conference on Neural Information Processing Systems (NIPS'01)*, pp. 817–823, Vancouver, British Columbia, Canada.

Kearns, M., Littman, M., & Singh, S. (2001b). Graphical models for game theory. In *Proc. of the 17th International Conference on Uncertainty in AI (UAI'01)*, pp. 253–260, Seattle, Washington, USA.

Kearns, M., & Mansour, Y. (2002). Efficient nash computation in large population games with bounded influence. In *Proc. of the 18th International Conference on Uncertainty in AI (UAI'02)*, pp. 259–266, Edmonton, Alberta, Canada.

Koller, D., & Megiddo, N. (1992). The complexity of two-person zero-sum games in extensive form. *Games and Economic Behavior, 2*, 528–552.

Koller, D., & Megiddo, N. (1996). Finding mixed strategies with small supports in extensive form games. *International Journal of Game Theory*, *14*, 73–92.

Koller, D., Megiddo, N., & von Stengel, B. (1996). Efficient computation of equilibria for extensive two-person games. *Games and Economic Behavior*, *14*, 220–246.

Koller, D., & Milch, B. (2001). Multi-agent influence diagrams for representing and solving games. In *Proc. of the 7th International Joint Conference on Artificial Intelligence (IJCAI'01)*, pp. 1027–1034, Seattle, Washington, USA.

Leyton-Brown, K., & Tennenholtz, M. (2003). Local-effect games. In *Proc. of the 18th International Joint Conference on Artificial Intelligence (IJCAI'03)*, pp. 772–780, Acapulco, Mexico.

Maier, D. (1986). *The Theory of Relational Databases*, Rochville, Md, Computer Science Press.

Mas-Colell, A., Whinston, M., & Green, J. (1995). *Microeconomic Theor*. Oxford University Press.

Maskin, E. (1985). The theory of implementation in nash equilibrium. *Social Goals and Organization: Essays in memory of Elisha Pazner*, 173–204.

McKelvey, R., & McLennan, A. (1996). Computation of equilibria in finite games. *Handbook of Computational Economics*, 87–142.

Megiddo, N., & Papadimitriou, C. (1991). On total functions, existence theorems, and computational complexity. *Theoretical Computer Science*, *81(2)*, 317–324.

Monderer, D., & Shapley, L. (1993). Potential games. *Games and Economic Behavior*.

Nash, J. (1951). Non-cooperative games. *Annals of Mathematics*, *54(2)*, 286–295.

Osborne, M., & Rubinstein, A. (1994). *A Course in Game Theory*. MIT Press.

Owen, G. (1982). *Game Theory*. Academic Press, New York.

Papadimitriou, C. (1994a). *Computational Complexity*. AAddison-Wesley, Reading, Mass.

Papadimitriou, C. (1994b). On the complexity of the parity argument and other inefficient proofs of existence. *Journal of Computer and System Sciences*, *48(3)*, 498–532.

Papadimitriou, C. (2001). Algorithms, games, and the internet. In *Proc. of the 28th International Colloqium on Automata, Languages and Programming (ICALP'01)*, pp. 1–3, Crete, Greece.

Robertson, N., & Seymour, P. (1986). Graph minors ii. algorithmic aspects of tree width. *Journal of Algorithms*, *7*, 309–322.

Rosenthal, R. (1973). A class of games possessing pure-strategy nash equilibria. *International Journal of Game Theory*, *2*, 65–67.

Schoenebeck, G.R., & Vadhan, S.P. (2005). The Computational Complexity of Nash Equilibria in Concisely Represented Games. *Electronic Colloquium on Computational Complexity, Report TR05-052.*

Stockmeyer, L., & Meyer, A. (1973). Word problems requiring exponential time: Preliminary report. In *Proc. of the 5th Annual ACM Symposium on Theory of Computing (STOC'73)*, pp. 1–9.

Vardi, M. (2000). Constraint satisfaction and database theory: a tutorial. In *Proc. of the 19th ACM SIGMOD-SIGACT-SIGART Symposium on Principles of Database Systems*, pp. 76–85, Dallas, Texas, USA.

Venkateswaran, H. (1991). Properties that characterize logcfl. *Journal of Computer and System Sciences, 43(2)*, 380–404.

Vickrey, D. amd Koller, D. (2002). Multi-agent algortihms for solving graphical games. In *Proc. of the 18th National Conference on Artificial Intelligence (AAAI'02)*, p. 345251 Edmonton, Alberta, Canada.

Yannakakis, M. (1981). Algorithms for acyclic database schemes. In *Proc. of the 7th International Conference on Very Large Data Bases (VLDB81)*, p. 8294 Cannes, France.